\documentclass[traditabstract]{aa} 
\usepackage{txfonts}
\usepackage{graphicx}
\usepackage{natbib}
\usepackage{color}

\newcommand{\hyeri}{\mbox{\object{HY\,Eri}}}
\newcommand{\chisq}{$\chi^2$}

\newcommand{\lyalpha}{Ly$\alpha$}
\newcommand{\halpha}{H$\alpha$}
\newcommand{\hbeta}{H$\beta$}
\newcommand{\hgamma}{H$\gamma$}
\newcommand{\hdelta}{H$\delta$}

\newcommand{\heii}{HeII$\lambda$4686}
\newcommand{\porb}{$P_\mathrm{orb}$}
\newcommand{\kms}{km\,s$^{-1}$}

\newcommand{\teff}{$T_\mathrm{eff}$}
\newcommand{\erg}{erg\,s$^{-1}$}
\newcommand{\ergs}{erg\,cm$^{-2}$s$^{-1}$}

\newcommand{\msol}{$M_\odot$\,yr$^{-1}$}
\newcommand{\cmkpc}{cm\,kpc$^{-1}$}
\newcommand{\ten}[2]{#1\times 10^{#2}}

\newcommand{\rsun}{$R_\odot$}
\newcommand{\msun}{$M_\odot$}

\newcommand{\rwd}{$R_\mathrm{wd}$}

\begin{document}

\title{Neglected X-ray discovered polars: II: The peculiar \\ eclipsing binary HY~Eridani}

\author{
Beuermann, K. \inst{1} \and 
Burwitz, V. \inst{2} \and 
Reinsch, K. \inst{1} \and 1
Schwope, A. \inst{3} \and
Thomas, H.-C. \inst{4}\thanks{Deceased 18 January 2012}
} 

\institute{ 
  Institut f\"ur Astrophysik, Georg-August-Universit\"at,
  Friedrich-Hund-Platz 1, D-37077 G\"ottingen, Germany \and
  MPI f\"ur extraterrestrische Physik, Giessenbachstr. 6, 85740
  Garching, Germany \and
  Leibniz-Institut f\"ur Astrophysik Potsdam (AIP), An der Sternwarte
  16, 14482 Potsdam, Germany \and
  MPI f\"ur Astrophysik, Karl-Schwarzschild-Str. 1, D-85740 Garching,
  Germany
}

\date{Received 3 September 2019; accepted 2 January 2020}

\authorrunning{K. Beuermann et al.} 
\titlerunning{Neglected X-ray discovered polars: II.}

\abstract{We report on the X-ray observations of the eclipsing polar
  HY\,Eri (RX\,J0501--0359), along with its photometric,
  spectrophotometric, and spectropolarimetric optical variations,
  collected over 30 years. With an orbital period of 2.855\,h, \hyeri\
  falls near the upper edge of the $2\!-\!3$~h period gap. After 2011,
  the system went into a prolonged low state, continuing to accrete at
  a low level. We present an accurate alias-free long-term orbital
  ephemeris and report a highly significant period change by 10\,ms
  that took place over the time interval from 2011 to 2018. We
  acquired a high-quality eclipse spectrum that shows the secondary
  star as a dM5--6 dwarf at a distance
  $d\!=\!1050\!\pm\!110$\,pc. Based on phase-resolved cyclotron and
  Zeeman spectroscopy, we identify the white dwarf (WD) in \hyeri\ as
  a two-pole accretor with nearly opposite accretion spots of 28 and
  30\,MG. The Zeeman analysis of the low state spectrum reveals a
  complex magnetic field structure, which we fit by a multipole
  model. We detected narrow emission lines from the irradiated face of
  the secondary star, of which Mg\,I$\lambda5170$ with a radial
  velocity amplitude of $K_2'\!=\!139\pm10$\,\kms (90\% confidence)
  tracks the secondary more reliably than the narrow \halpha\
  line. Based on the combined dynamical analysis and spectroscopic
  measurement of the angular radius of the WD, we obtain a primary
  mass of $M_1\!=\!0.42\,\pm\,0.05$\,\msun\ (90\% confidence errors),
  identifying it as a probable He WD or hybrid HeCO WD.  The secondary
  is a main sequence star of $M_2\!=\!0.24\,\pm\,0.04$\,\msun\ that
  seems to be slightly inflated. The large distance of \hyeri\ and the
  lack of similar systems suggest a very low space density of polars
  with low-mass primary. According to current theory, these systems
  are destroyed by induced runaway mass transfer, suggesting that
  \hyeri\ may be doomed to destruction. Over the last 30 years,
  \hyeri\ experienced high and low states with mass transfer rates
  that differed by three orders of magnitude, varying between
  $\dot{M}\!\simeq\!10^{-9}$\,\msol and $10^{-12}$\,\msol. At a
  galactic latitude of -26.1\degr, \mbox{it is located about 500\,pc below
  the galactic plane}.}

\keywords{stars: cataclysmic variables -- stars: magnetic fields --
  stars: binaries: close -- stars: binaries: eclipsing -- stars: white
  dwarf -- stars: individual: HY Eri}

\maketitle

\section{Introduction}

Of the more than 1200 cataclysmic variables in the final 2016 edition of
the \citet{ritterkolb03} catalog, 114 are confirmed polars (or AM
Herculis binaries), which contain a late-type main sequence star and
an accreting magnetic white dwarf in synchronous rotation. The name
``polar'' was coined by \citet{krzeminskiserkowski77} to describe the
high degree of circular polarization, which became one of the
hallmarks of the class. Another is the large portion of the
bolometric luminosity emitted in high states of accretion in form of
soft and hard X-ray emission, which led to the discovery of the
majority of the known systems. Many individual polars are
characterized by idiosyncrasies, which distinguish them from their
peers and provide special insight into the physics of
polars. Unresolved questions relate, for example, to the physics of accretion
\citep{bonnetbidaudetal00,busschaertetal15,bonnetbidaudetal15},
various aspects of close-binary evolution
\citep{webbinkwickramasinghe02,liebertetal05,kniggeetal11}, and the
generation and structure of the magnetic field of the white dwarf (WD)
\citep{beuermannetal07,wickramasingheetal14,ferrarioetal15}.

Our optical programs for identifying high-galactic latitude ROSAT
X-ray sources \citep{thomasetal98,beuermannetal99,schwopeetal02} have
led to the discovery of 27 new polars. Twenty sources have been
described in previous publications. In this series of three papers, we
present results on the remaining seven. Paper~I
\citep{beuermannetal17} describes V358\,Aqr (=\,RX\,J2316--05), a
system that experiences giant flares on its secondary star. Here we
present a comprehensive analysis of the eclipsing polar \hyeri\
(=\,RX\,J0501--03) based on data collected over three decades. Our
early conference paper \citep{burwitzetal99} represents the only
previous account of the system in the literature. The third paper of
this series will contain shorter analyses of RX\,J0154$\!-\!59$,
RX\,J0600$\!-\!27$, RX\,J0859+05, RX\,J0953+14, and
RX\,J1002$\!-\!19$, of which three have not been addressed previously
either.

\section{Observations}
\label{sec:obs}

\subsection{X-ray data}
\label{sec:obsx}

\hyeri, located at RA(2000)\,=\,$05^\mathrm{h}01^\mathrm{m}46\farcs4$,
DEC(2000)\,= $-03\degr59\arcmin20\arcsec$ ($l,b\!=\!203.5,-26.1$) was
discovered as a very soft X-ray source in the RASS
\citep{bolleretal16}\footnote{http://cdsarc.u-strasbg.fr/viz-bin/qcat?J/A+A/588/A103}
and identified by us spectroscopically with an eclipsing polar
\citep{beuermannthomas93,beuermannetal99,burwitzetal99}.  Follow-up
pointed ROSAT observations were performed 1992 and 1993 with the
Position Sensitive Proportional Counter (PSPC) as the detector and 1995
and 1996 with the High Resolution Imager (HRI). These data were
originally published by \citet{burwitzetal99} and reanalyzed for the
present study. We also analyzed the previously unpublished data
taken in 2002 with XMM-Newton equipped with the EPIC camera.  On all
occasions, HY Eri was encountered with an X-ray flux that corresponds
to a high or near high state\footnote{ Photometrically, "high" and
  "low" states refer to two brightness levels between which polars
  oscillate in their long-term light curves. Spectroscopically, a
  "high" state is usually characterized by intense He\,II$\lambda4686$
  line emission, which is absent in a "low" state. Physically, "high"
  refers to accretion rates adequate to drive the standard secular
  evolution of CVs, while in a "low" state, accretion ceases either
  completely or is reduced to a trickle. "Intermediate" refers to
  temporary states in between.} (Table~\ref{tab:xray}).

\begin{table}[t]
\begin{flushleft}
\caption{Time-resolved X-ray observations of \hyeri}
\begin{tabular}{l@{\hspace{3.0mm}}l@{\hspace{2.0mm}}l@{\hspace{2.0mm}}r@{\hspace{3.0mm}}c@{\hspace{1.0mm}}c}\\[-5ex]
 \hline\hline \\[-1.0ex]
~Dates                 & Instrument &  Band         & Total & State~~ & Ref. \\
                       &            &  (keV)        & (ks) &       &      \\[0.5ex]
  \hline\\[-1ex]                                                             
24--26\,Aug\,1990      & RASS PSPC  &  $0.1\!-\!2$  &  0.5 & high  &  (1) \\
24\,Feb\,1992          & ROSAT PSPC &  $0.1\!-\!2$  &  2.5 & high  &  (1) \\
15--22\,Feb\,1993      & ROSAT PSPC &  $0.1\!-\!2$  &  1.9 & high  &  (1) \\
 8--16\,Sep\,1995      & ROSAT HRI  &  $0.1\!-\!2$  & 16.5 & high  &  (1) \\
26\,Feb--19\,Mar\,1996 & ROSAT HRI  &  $0.1\!-\!2$  & 28.8 & high  &  (1) \\
24\,Mar\,2002          & XMM MOS+pn &  $0.2\!-\!10$ &  6.6 & high  &  (2) \\[1.0ex]
\hline\\
\end{tabular}\\[-1.0ex]
\footnotesize{(1) Burwitz et al (1999), (2) This work}
\label{tab:xray}
\end{flushleft}

\vspace{-4mm}
\end{table}

\begin{table}[t]
\begin{flushleft}
\caption{Journal of time-resolved optical photometry}
\begin{tabular}{l@{\hspace{0.0mm}}c@{\hspace{0.0mm}}c@{\hspace{1.0mm}}c@{\hspace{2.0mm}}c@{\hspace{3.0mm}}c@{\hspace{2.0mm}}c}\\[-5ex]
  \hline\hline \\[-1.0ex]
Dates         & Number & Band & Expos.&  Total & State & Tel. \\
              & nights &      & (s)     &   (h)  &  &      \\[0.5ex]
  \hline\\[-1ex]                                                
Feb 1994, Jan 1996  &                3 &  V        & 20/60 & \hspace{-1.5mm}11.7 & high    & (1) \\        
3 -- 5 Feb 2001     &                2 & R,Gunn\,i & 60    & 1.1                 & high    & (2) \\
17 -- 22 Jan 2010   &                5 &  WL       & 10/60 & \hspace{-1.5mm}10.6 & high    & (3) \\
 8 -- 18 Nov 2010   &                3 &  WL       & 10/60 &                 4.2 & interm. & (3) \\
Feb -- Oct2011      &                3 &  WL       & 10/60 &                 3.6 & interm. & (3) \\
Aug 2014 -- Jan 2015&\hspace{-1.5mm}10 &  WL       & 15/60 &                 8.0 & low     & (3) \\
24/27 Oct 2016      &                2 & grizJHK   & 35    &                 5.1 & interm. & (4) \\
Sep 2017 -- Jan 2019&\hspace{-1.5mm}23 &  WL       & 15/60 & \hspace{-1.5mm}34.8 & low     & (5) \\
Nov 2017            &                2 & grizJHK   & 35    &                 4.7 & low     & (4)\\
Feb 2018            &                1 & grizJHK   & 35    &                 3.0 & low     & (4)\\[1.0ex]  
 \hline\\
\end{tabular}\\[-1.0ex]
\footnotesize{
 (1) ESO\,/\,Dutch 0.9\,m, (2) ESO 3.6\,m with EFOSC2, (3) McD, MONET\,/\,N 1.2\,m, (4) MPG/ESO 2.2\,m with GROND, (5) SAAO, MONET\,/\,S 1.2\,m}
\label{tab:phot}
\end{flushleft}

\vspace{-3mm}
\end{table}

\begin{table}[t]
\begin{flushleft}
\caption{Time-resolved spectroscopy and circular spectropolarimetry}
\begin{tabular}{@{\hspace{1.0mm}}l@{\hspace{0.0mm}}c@{\hspace{1.0mm}}c@{\hspace{1.0mm}}c@{\hspace{1.0mm}}c@{\hspace{1.0mm}}c@{\hspace{2.0mm}}c@{\hspace{2.0mm}}c}\\[-5ex]
\hline\hline \\[-1.0ex]
Dates          &   Band     & Resol. &  Num- & Expos. & Total & State&Instr.\\
               &   (\AA)    & (\AA)  &  ber  &  (min) &  (h)  &       &     \\[0.5ex]
\hline\\[-1ex]                                                             
13-17\,Dec\,1993 & 3500--9500 &              25 &              50 &              10.0 &              10.6 & high & (1)\\
14\,Nov\,1995    & 3800--9119 &              25 &              24 & \hspace{1.0mm}2.0 & \hspace{1.0mm}1.1 & high & (2)\\
15\,Nov\,1995    & 3869--5109 & \hspace{1.0mm}8 &              17 &              10.0 & \hspace{1.0mm}3.2 & high & (2)\\
20\,Nov\,2000    & 6340--10400&              10 & \hspace{1.0mm}8 & \hspace{1.0mm}0.5 & \hspace{1.0mm}0.5 & high & (3)  \\
20\,Nov\,2000    & 6340--10400&              10 & \hspace{1.0mm}1 &              10.0 &                   &      & (3)\\
31\,Dec\,2008    & 3800--9200 &              10 &              40 & \hspace{1.0mm}6.5 & \hspace{1.0mm}5.6 & low  & (4)\\[1.0ex]
\hline\\
\end{tabular}\\[-1.0ex]
\footnotesize{(1) ESO 1.5\,m, B\&C spectrograph, (2) MPG/ESO 2.2\,m with EFOSC2,
(3)~ESO VLT UT1 with FORS1, (4) ESO VLT UT2 with FORS1}
\label{tab:spec}
\end{flushleft}

\vspace{-3mm}
\end{table}

\subsection{Optical photometry}
\label{sec:phot}

Orbital $BVRI$ light curves and $V$-band eclipse light curves were
measured in 1994 and 1996 with the ESO/Dutch 0.9\,m telescope
\citep{burwitzetal99}.  Extensive white-light (WL) photometry,
performed between 2010 and 2019 with the 1.2\,m MONET/N and MONET/S
telescopes at the McDonald Observatory and the South African
Astrophysical Observatory, respectively, allowed us to establish an
alias-free long-term ephemeris. Seven-color $grizJHK$ photometry was
performed with the MPG/ESO~2.2\,m telescope equipped with the
GROND\footnote{Gamma-ray Burst Optical/Near-infrared Detector.}
photometer in 2016, 2017, and 2018.  A log of the observations is
provided in Table~\ref{tab:phot}.
We measured magnitudes relative to the dM1-2 star
SDSS\,050146.02-040042.2 (referred to as C1), which is located
43\arcsec\ E and 4\arcsec\ N of \hyeri\ and has Sloan AB magnitudes
$g\!=\!16.91, r\!=\!16.38, i\!=\!16.23$, and $z\!=\!16.17$. Its color,
$g\!-\!i\!=\!0.68$, is similar to the low-state Sloan color of \hyeri,
$g\!-\!i\!=\!0.71$.  \hyeri\ is separated by only 6\,\farcs 2 from the
center of a galaxy with Sloan $r\!=\!18.44$.  All accepted eclipse
light curves were taken in sufficiently good seeing to escape
spillover from the galaxy.

\begin{figure*}[t]
\includegraphics[height=89.0mm,angle=270,clip]{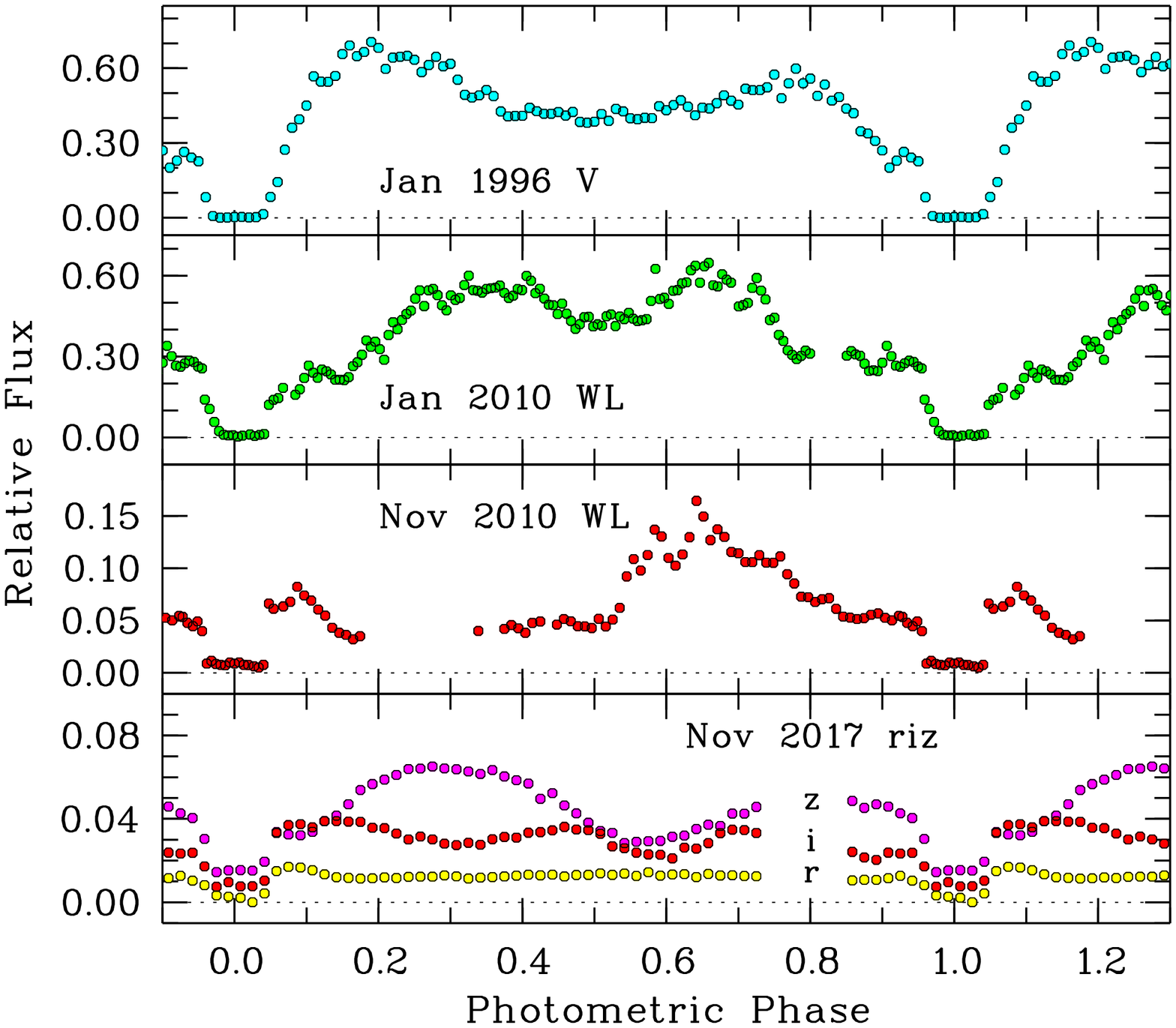}
\hfill
\includegraphics[height=89.0mm,angle=270,clip]{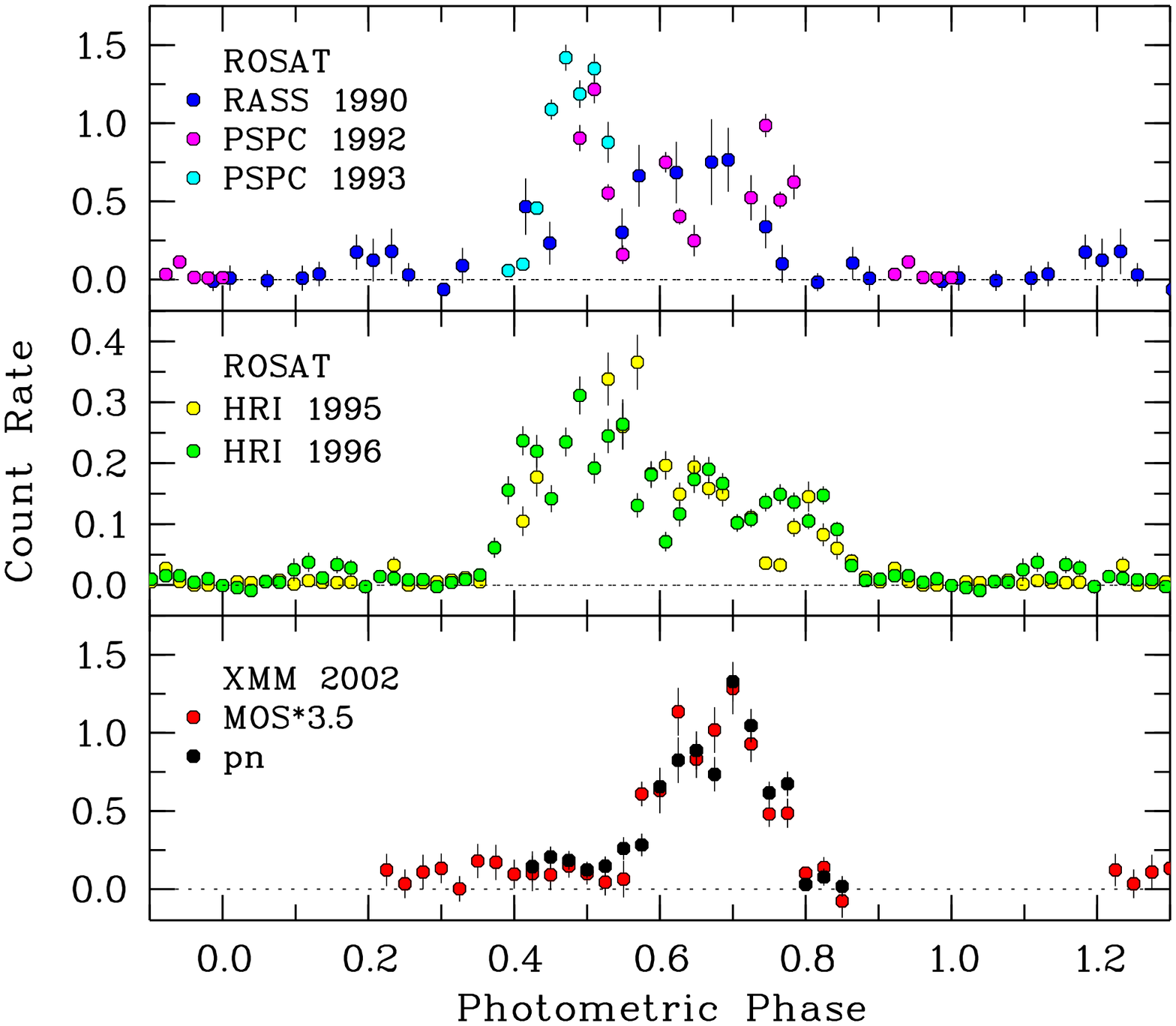}

\caption[chart]{\emph{Left, from top: } Orbital light curves of
  \hyeri\ in the 1996 and January 2010 high states, November 2010
  intermediate, and 2017 low state, binned to $\sim\!2$\,min time
  resolution.  \emph{Right, from top: } Binned ROSAT X-ray light
  curves measured with either the PSPC or the HRI as detectors and
  XMM-Newton light curve measured with the MOS and pn detectors of the
  EPIC camera.}
\label{fig:lc}
\end{figure*}

\subsection{Optical spectroscopy}
\label{sec:obsspec}

Follow-up time-resolved spectrophotometric observations of \hyeri\ in
its high state were performed in 1993 and 1995, using the ESO 1.5\,m
telescope with the Boller \& Chivens spectrograph and the ESO/MPG 2.2
m telescope with EFOSC2, respectively. In the latter run, grisms G1
and G3 yielded low- and medium-resolution spectra with FWHM
resolutions of 25\,\AA\ and 8\,\AA\ that covered the entire optical
band and the blue band, respectively. Using the photometrically
established eclipse ephemeris, a 10\,min exposure in the near-total
eclipse was taken with the ESO/VLT UT1 equipped with FORS\,1 on 20
November 2000. Grating G300I provided coverage of the red part of the
spectrum, which is dominated by the secondary star of \hyeri. The
magnetic nature of the WD was studied in 2008 by phase-resolved
low-resolution circular spectropolarimetry performed with the ESO/VLT
UT2 and FORS1. Grism G300V provided coverage from
3800--9200\,\AA. Table~\ref{tab:spec} lists the wavelength ranges,
number of spectra, exposure times, and total times spent on source.

\begin{figure*}[t]
\includegraphics[height=89.0mm,angle=270,clip]{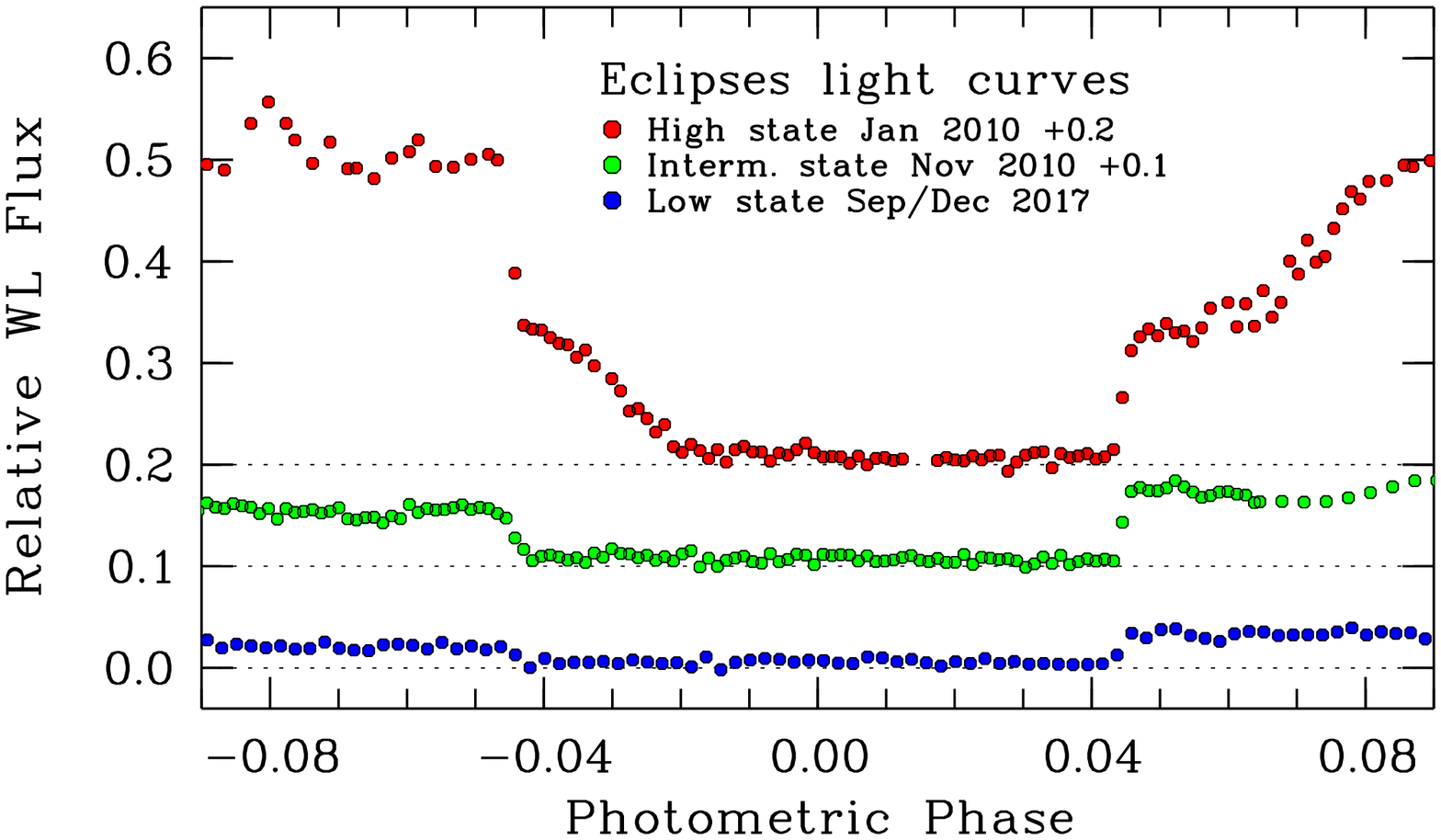}
\hfill
\includegraphics[height=91.0mm,angle=270,clip]{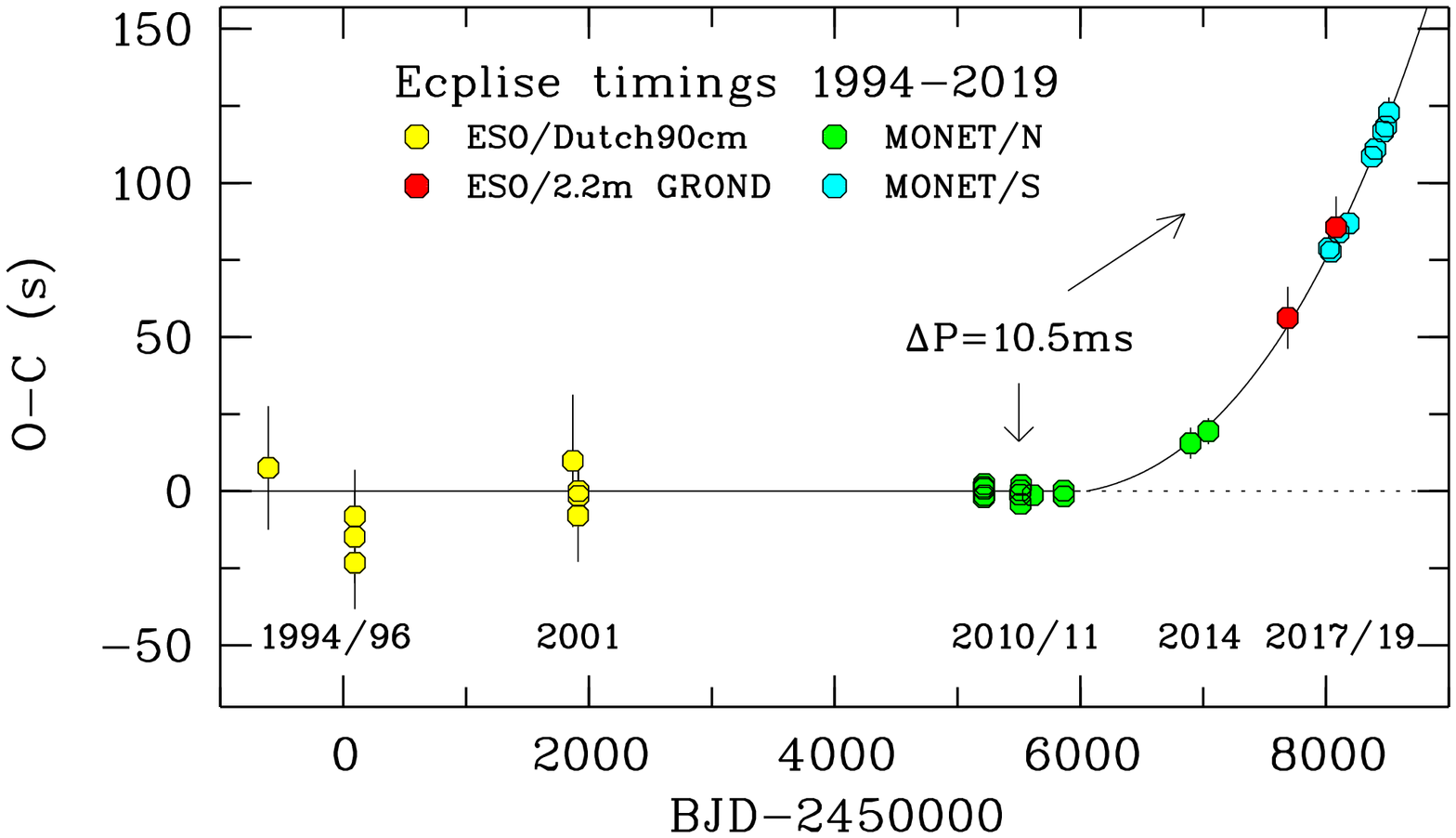}

\caption[chart]{\emph{Left: } Eclipse light curves in the high,
  intermediate, and in the low state, the former two shifted upward to
  avoid overlap. Phases are from Eq.~\ref{eq:ephem1} for
  the high and intermediate states and from and Eq.~\ref{eq:ephem2}
  for the low state. \emph{Right: } $O-C$ diagram for the deviations
  of the mid-eclipse times from the linear ephemeris of
  Eq.~\ref{eq:ephem1}, showing the change in orbital period. }
\label{fig:ecl}

\vspace{-2mm}
\end{figure*}

\subsection{Synthetic white light photometry}
\label{sec:obswl}

As described in Paper\,I, we performed synthetic photometry in order
to tie the WL measurements obtained with the MONET telescopes into the
standard $ugriz$ system. We defined a MONET-specific WL AB
magnitude $w$, which has its pivot wavelength
$\lambda_\mathrm{piv}\!=\!6379$\,\AA\ in the red part of the Sloan $r$
band. For a star with the colors of comparison star C1, the color
difference is $w\!-\!r\!=\!+0.07$. Hence, unity relative WL flux
corresponds to $r\!=\!16.38$ and $w\!=\!16.45$.  For WL measurements
of other stars, $w$ is a measured quantity and is related to Sloan $r$
by $r\!=\!w-(w\!-\!r)_\mathrm{syn}$. For a wide range of incident
spectra, the synthetic color $|(w\!-\!r)_\mathrm{syn}|\!\lesssim\!0.1$.
Hence, $w\!\simeq\!r$ is typically correct within 0.1 mag, except for
very red stars.

\section{Optical light curves}
\label{sec:res}

\subsection{Orbital  light curves}
\label{sec:olc}

In Fig.~\ref{fig:lc}, we show the $V$-band and WL light curves in the
high states of 11 January 1996 and 19 January 2010, respectively, the
WL light curve in the intermediate state of 14-18 November 2010, and
the $riz$ low-state light curves of 24-25 November 2017. Orbital phase
$\phi\!=\!0$ is defined by the eclipse ephemeris provided in
Sect.~\ref{sec:ephem}.  \hyeri\ reached orbital maxima of $V\!=\!16.8$
and $w\!=\!17.1$ in the high states and $w\!=\!18.7$ in the
intermediate state. The 2017 peak magnitudes were $z\!=\!19.1$ and
$i\!=\!19.7$, while $r$ stayed at $21$ throughout the orbit outside
eclipse. In all states, the light curves exhibit the signatures of
emission from two accretion regions, being shaped by cyclotron
beaming. The same holds for the light curves in the right panel of
Fig.~\ref{fig:meanspec}. Borrowing from the insight provided by the
low-state spectropolarimetry (Sect.~\ref{sec:cyc}), we identify, for
instance, the double-humped $z$-band light curve in the lower left
panel of Fig.~\ref{fig:lc} as cyclotron emission in the 4th harmonic
from two accretion regions best viewed at phases
$\varphi\!\simeq\!0.35$ and $0.85$.  The light curves in WL are less
easily interpreted because of the lack of wavelength resolution.  We
loosely refer to the emission regions best seen around $\phi\!=\!0.85$
and $\phi\!\simeq\!0.35$ as ``pole\,1" and ``pole\,2" or ``spot\,1''
and ``spot\,2'', respectively. In the high state, the emissions from
both poles become an inextricable conglomerate.  The available
evidence suggests that \hyeri\ is a permanent two-pole emitter.

\subsection{Eclipse light curves}
\label{sec:eclipse}

We collected a total of 41 eclipses of the hot spot on the WD by the
secondary star, 13 in various high states, eight in intermediate
states, and 20 in low states. Ingress and egress of the hot spot take
place in typically less than 20\,s.  In Fig~\ref{fig:ecl}, we show the
eclipse light curves in the three states at the original time
resolutions of 13\,s (exposure and readout) for the high and
intermediate states and and 22\,s for the low state. A characteristic
feature of the high state is the delayed eclipse of the accretion
stream. This component disappears, when the accretion rate drops.

\begin{table}[tb]
\begin{flushleft}
\caption{Observed mid-eclipse times of \hyeri\ }
%\medskip
\begin{tabular}{@{\hspace{0.0mm}}r@{\hspace{2.0mm}}c@{\hspace{2.0mm}}c@{\hspace{2.0mm}}r@{\hspace{2.0mm}}r@{\hspace{3.0mm}}c@{\hspace{1.0mm}}c@{\hspace{1.0mm}}c@{\hspace{1.0mm}}c}\\[-5ex]
\hline\hline \\[-1.5ex]
Cycle & BJD(TDB)    & n &Error&$O\!-\!C$ &  State  & Expos&Band& Instr.\\
      & 2400000+    &   &(s)\hspace{1mm}  & (s)\hspace{1mm}       &         & (s)  &    &     \\[0.5ex]    
\hline\\[-1ex]                                                              
-54428 & 49390.622004 & 1 & 20.0 &   7.6 & high    & 20 & V    & (1) \\       
-48518 & 50093.728987 & 1 & 15.0 & -14.8 & high    & 60 & V    & (1) \\       
-48511 & 50094.561847 & 1 & 15.0 &  -8.2 & high    & 30 & V    & (1) \\       
-48510 & 50094.680642 & 1 & 15.0 & -23.2 & high    & 30 & V    & (1) \\       
-33598 & 51868.747892 & 1 & 21.5 &   9.8 & high    & 60 & Spec & (2) \\         
-33229 & 51912.647277 & 1 & 15.0 &  -7.9 & high    & 60 & R    & (3) \\         
-33213 & 51914.550874 & 1 & 10.0 &   0.1 & high    & 60 & R    & (3) \\         
-33212 & 51914.669824 & 1 & 10.0 &  -1.5 & high    & 60&Gunn\,i& (3) \\         
 -5483 & 55213.563377 & 1 &  2.1 &   1.3 & high    & 10 & WL   & (4) \\         
 -5474 & 55214.634061 & 1 &  3.2 &  -2.0 & high    & 10 & WL   & (4) \\         
 -5465 & 55215.704814 & 1 &  1.2 &   0.8 & high    & 10 & WL   & (4) \\         
 -5449 & 55217.608337 & 1 &  1.2 &   2.3 & high    & 10 & WL   & (4) \\         
 -5439 & 55218.797986 & 1 &  2.6 &  -1.4 & high    & 10 & WL   & (4) \\         
 -3000 & 55508.963566 & 1 &  2.2 &  -1.1 & interm. & 10 & WL   & (4) \\         
 -2959 & 55513.841263 & 1 &  1.8 &  -4.1 & interm. & 10 & WL   & (4) \\         
 -2958 & 55513.960283 & 1 &  1.2 &   0.3 & interm. & 10 & WL   & (4) \\         
 -2916 & 55518.957004 & 1 &  1.5 &   2.0 & interm. & 10 & WL   & (4) \\         
 -2129 & 55612.585630 & 1 &  2.8 &  -1.3 & interm. & 10 & WL   & (4) \\         
   -17 & 55863.848336 & 1 &  2.1 &   0.2 & interm. & 10 & WL   & (4) \\         
     0 & 55865.870789 & 1 &  1.1 &  -1.7 & interm. & 10 & WL   & (4) \\         
  8675 & 56897.927726 & 3 &  5.0 &  15.6 & low     & 20 & WL   & (4) \\         
  9892 & 57042.713137 & 1 &  4.2 &  19.5 & low     & 15 & WL   & (4) \\         
 15315 & 57687.882864 & 1 & 10.0 &  56.2 & interm  & 33 & iz   & (5) \\         
 18137 & 58023.613860 & 1 &  1.7 &  78.9 & low     & 15 & WL   & (6) \\         
 18279 & 58040.507455 & 1 &  2.8 &  77.7 & low     & 15 & WL   & (6) \\         
 18635 & 58082.860538 & 1 & 10.0 &  85.6 & low     & 33 & iz   & (5) \\         
 18800 & 58102.490416 & 1 &  4.6 &  84.0 & low     & 15 & WL   & (6) \\         
 19496 & 58185.292928 & 2 &  3.8 &  86.9 & low     & 15 & WL   & (6) \\       
 21070 & 58372.550511 & 4 &  3.0 & 109.1 & low     & 15 & WL   & (6) \\       
 21322 & 58402.530740 & 3 &  4.2 & 111.0 & low     & 15 & WL   & (6) \\       
 21835 & 58463.561939 & 2 &  2.9 & 116.4 & low     & 15 & WL   & (6) \\       
 22237 & 58511.387633 & 1 &  6.9 & 127.2 & low     & 15 & WL   & (6) \\[1.0ex]                                          
 \hline\\                                                                       
\end{tabular}\\[-1.0ex]
\footnotesize{(1) ESO/Dutch 0.9\,m, (2) ESO VLT UT1, EFOSC\,1, (3) ESO 3.6\,m, EFOSC\,2, (4) McD MONET/N 1.2\,m, (5) MPG/ESO 2.2\,m with GROND, (6) SAAO MONET/S 1.2\,m}\\[-3.0ex]
\label{tab:ephem}                                                              
\end{flushleft}

\vspace{-0mm}
\end{table}

The eclipse was modeled by the occultation of a circular disk of
uniform surface brightness, which represents either the WD or the hot
spot on the WD. The parameters of the fit are the mid-eclipse time,
the FWHM, and the duration of ingress or egress.  We improved the
statistics and reduced the timing error in the low state, when the
star became as faint as $w\!=\!20.7$ outside eclipse, by fitting up to
$n\!=\!4$ eclipses on a barycentric time scale if taken nearby
in time. The resulting mid-eclipse times are listed in
Table~\ref{tab:ephem}. In the cases with $n\!>\!1$, the cycle number
of the best-defined eclipse is quoted. The measured FWHM of the
eclipse is the same for the different accretion states with a mean
value of $\Delta\,t_\mathrm{ecl}\!=\!910.6\!\pm\!1.5$\,s or
$\Delta\,\phi\!=\!0.08859\!\pm\!0.00015$ in phase units. This is the
longest relative eclipse width of all polars. We did not detect the
ingress and egress of the WD photosphere because our WL observations
are dominated by cyclotron emission and the measurements in the GROND
$gr$ filters lacked the required time resolution. The quoted
mid-eclipse times refer to the hot spot on the WD and may deviate from
true inferior conjunction of the secondary by up to $\sim\!30$\,s or
$\sim\!0.003$ in phase.

The relative WL flux in the totality is the same in the high,
intermediate, and low states, with a mean of $0.0060\pm0.0005$ or
$w\!=\!22.0\!\pm\!0.1$. Using a color correction
$w\!-\!r\!\simeq\!-0.3$, appropriate for the secondary star, we find
$r\!\simeq\!22.3$, which compares favorably with the result of our
spectrophotometry in eclipse reported Sect.~\ref{sec:vlt}. Hence, the
residual WL flux in eclipse is largely that of the secondary star.

\subsection{Eclipse ephemeris}
\label{sec:ephem}

We corrected the UTC eclipse times to Barycentric Dynamical Time
(TDB), using the tool provided by
\citet{eastmanetal10}\,\footnote{http://astroutils.astronomy.ohio-state.edu/time/},
which accounts also for the leap seconds.  The complete set of eclipse
times is presented in Table~\ref{tab:ephem}.  The 2010 and 2011 data and the
scanty earlier data define the  alias-free linear ephemeris
\begin{equation}
T_\mathrm{ecl}\!=\!\mathrm{BJD(TDB)}~~2455865.87081(1) + 0.118969076(2)\,E,~~
\label{eq:ephem1}
\end{equation}
with $\chi^2\!=\!19.6$ for 18 degrees of freedom (solid line, cycle
numbers $E\!\le\!0$). The $O\!-\!C$ offsets from the ephemeris of
Eq.~\ref{eq:ephem1} are displayed in Fig.~\ref{fig:ecl}, right
panel. This ephemeris became increasingly invalid after 2011 and the
more recent data are well represented by a cubic ephemeris for
$E\!>\!-6000$ (solid curve). Currently, $O\!-\!C$ relative to the
linear ephemeris of Eq.~(1) has exceeded 2\,min, which is much too
large to be explained by wandering of the accretion spot. The
mid-eclipse times of 2017 to 2019 follow the linear ephemeris
\begin{equation}
  T_\mathrm{ecl}\!=\!\mathrm{BJD(TDB)}~2455865.86951(14) + 0.118969198(8)\,E,~~~
\label{eq:ephem2}
\end{equation}
implying a period change relative to Eq.~(1) of
$10.5\!\pm\!0.7$\,ms. The mean rate of the period variation between
2011 and 2018 is $\dot P\!\simeq\!\ten{5}{-11}$\,ss$^{-1}$. The
period change started approximately, when the system entered a
prolonged low state after 2011. This is likely a coincidence, however,
because it was also in a low state in 2008 and temporarily in an
intermediate state in 2016.

\begin{table}[b]
\begin{flushleft}
\caption{Parameters of X-ray spectral fits for \hyeri}

\begin{tabular}{l@{\hspace{2.0mm}}l@{\hspace{3.0mm}}c@{\hspace{3.0mm}}c@{\hspace{1.0mm}}c@{\hspace{1.0mm}}c}\\[-3.0ex]
\hline\hline \\[-1.0ex]
Observation    & Detector & $T_\mathrm{bb}$ & $N_\mathrm{H}$ &\hspace{4.0mm} $F_\mathrm{bb}$ & $F_\mathrm{th}$ \\
               &          &  (eV)    & ($10^{20}$/cm$^2$s) &\multicolumn{2}{r}{($10^{-11}$\,erg/cm$^2$s)}  \\[0.5ex]
\hline\\[-1ex]                                                     
RASS 1990      & PSPC& \hspace{1.0mm}30 : & 3.0 fixed       &\hspace{4.0mm}9 :           &       \\
ROSAT 1992/93  & PSPC     & $37\!\pm\!6$  & $8.0\!\pm\!2.0$ &\hspace{2.0mm}$43\!\pm\!30$ &       \\
               &          & $44\!\pm\!3$  & 6.0 fixed       &\hspace{2.3mm}13.0          &       \\
               &          & $50\!\pm\!3$  & 3.0 fixed       &\hspace{4.0mm}1.7           &       \\
XMM 2002&\hspace{-2.0mm}MOS1+pn& $45\!\pm\!6$&$1.4\,+\!2.0,-\!1.4$ &\hspace{4.0mm}0.5    & 0.08  \\
               &          & $40\!\pm\!2$  & 3.0 fixed       &\hspace{4.0mm}1.5           & 0.09  \\[1.0ex]
\hline\\
\end{tabular}\\[-1.0ex]
\label{tab:xspec}
\end{flushleft}
\end{table}

\section{X-ray light curves and spectra}
\label{sec:xray}

The ROSAT soft X-ray light curves taken between 1990 and 1996
\citep{burwitzetal99} show a structured bright phase that extends from
$\phi\!\simeq\!0.4$ to $\phi\!\simeq\!0.8$ with low-level emission
over the remainder of the orbit. Binned versions of these data are
shown in Fig.~\ref{fig:lc}, right panels. We also included the
previously unpublished light curve measured with XMM-Newton and the
MOS and pn detectors of the EPIC camera in 2002. The lower right panel
of Fig.~\ref{fig:lc} shows the mean count rates of the two MOS
detectors and the pn-detector, respectively, with the former adjusted
by a factor of 3.5 upward. The X-ray bright phases in the ROSAT and
XMM-Newton light curves show some similarity with the WL optical light curve
of November 2010, suggesting that the observed intense X-ray emission
originates from pole\,1, phase-modulated by rotation of the WD and
possibly by internal absorption. The low-level emission around
$\phi\!=\!0.3$ may stem from pole\,2. The bright phase reached count
rates around 1.0\,PSPC\,cts\,s$^{-1}$, 0.4\,HRI cts\,s$^{-1}$, and
1\,EPIC-pn cts\,s$^{-1}$, suggesting that \hyeri\ was in some form of
high or intermediate-to-high state during these observations.

\begin{figure}[t]
\includegraphics[height=89.0mm,angle=270,clip]{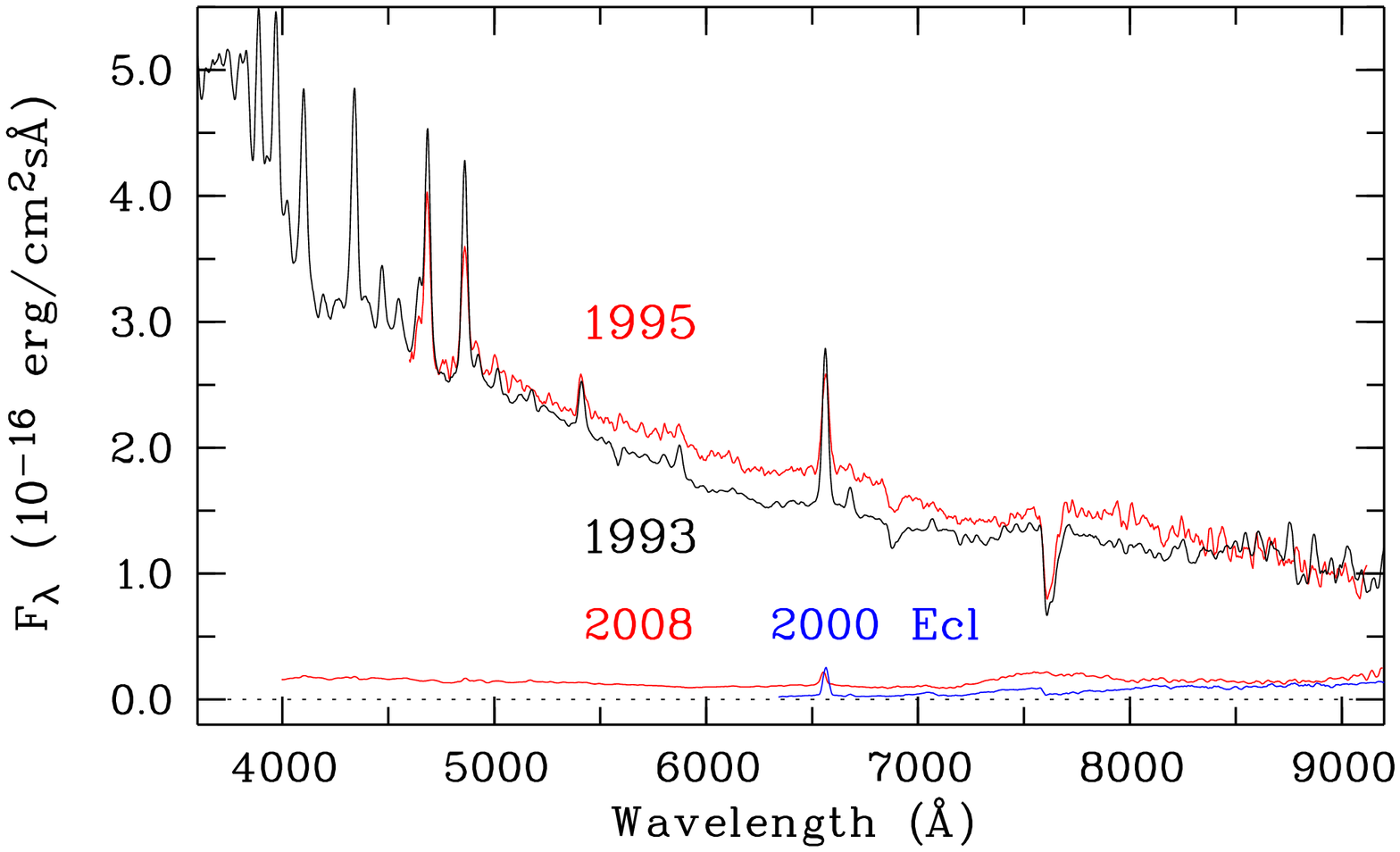}

\vspace{1.3mm}
\includegraphics[height=89.0mm,angle=270,clip]{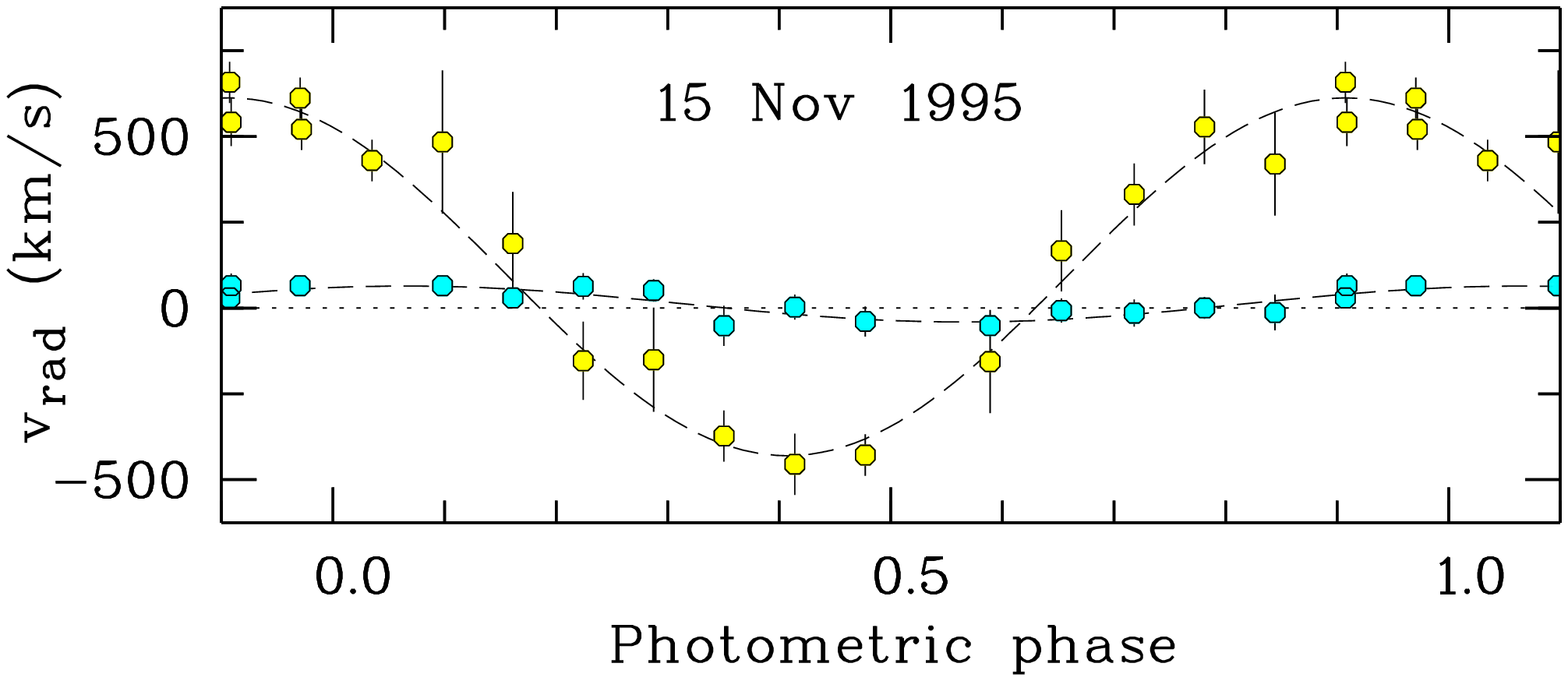}
\caption[chart]{\emph{Top: } Mean flux-calibrated low-resolution
  spectra of \hyeri\ in the high states of 1993 and 1995. For
  comparison, the eclipse spectrum of 2000 (blue curve) and the mean
  out-of-eclipse spectrum in the 2008 low state (red curve) are added
  on the same ordinate scale. \emph{Bottom: } Mean radial velocities
  of the broad emission lines of \hbeta\ and \heii\ (open circles) and
  of the line peaks (cyan dots) derived from medium-resolution spectra
  taken on 15 November 1995.}
\label{fig:spec}
\end{figure}

The ROSAT and XMM-Newton spectra (not shown) are characterized by an
intense soft X-ray and an underlying hard X-ray component, of which
the latter is well covered only in the XMM-Newton run. We fitted the
bright-phase spectra with the sum of a blackbody of temperature
k$T_\mathrm{bb}$ and a thermal component with the temperature fixed at
10\,keV, both attenuated by a column density $N_\mathrm{H}$ of cold
interstellar matter of solar composition\footnote{Using XSPEC version
  12.9.1n with TBabs*(bbody+apec)}. The fit parameters are listed in
Table~\ref{tab:xspec}. The values of k$T_\mathrm{bb}$ and
$N_\mathrm{H}$ differ substantially, indicating either true
variability or systematic uncertainties caused by the poor energy
resolution of the ROSAT PSPC, the lack of spectral coverage of the
XMM-Newton detectors below 0.2\,keV, or the inadequacy of fitting a
multi-temperature source by a single blackbody
\citep{beuermannetal12}. In any case, \hyeri\ is more strongly
absorbed than other polars. For the combined PSPC fit to the 1992 and
1993 data, $N_\mathrm{H}$ exceeds the total galactic column density
$N_\mathrm{H,gal}\!\simeq\!\ten{5.2}{20}$\,H-atoms\,cm$^{-2}$ at the
position of \hyeri\
\citep{hi4pi16}\,\footnote{http://vizier.u-strasbg.fr/viz-bin/VizieR?-source=J/A+A/594/A116}.
The galactic extinction at the position of \hyeri,
$A_\mathrm{V}\!=\!0.172$ \citep{schlegeletal98}, and the
$N_\mathrm{H}\!-\!A_\mathrm{V}$ conversion factor of
\citet{predehlschmitt95} yield
$N_\mathrm{H}\!\simeq\!\ten{3}{20}$\,H-atoms cm$^{-2}$, which
corresponds to the galactic dust layer. Forcing the fits to this
value, the ROSAT and XMM-Newton fits yield similar blackbody
temperatures and a bolometric soft X-ray flux of
$F_\mathrm{X}\!\simeq\!\ten{1.5}{-11}$\,\ergs.

\section{Optical spectroscopy and spectropolarimetry}
\label{sec:oss}

\subsection{High-state spectra}
\label{sec:spec}

In Fig.~\ref{fig:spec}, we show the mean low-resolution spectra taken on
13--17 December 1993 and 14 November 1995, when \hyeri\ was in high
states. They are characterized by a blue continuum, strong Balmer,
He\,I, and He\,II emission lines, the Balmer jump in emission, and
weak broad cyclotron lines at the red end. Phase-resolved radial
velocities were measured from medium-resolution blue spectra taken on
15 November 1995 (not shown). The Balmer and HeII emission lines were
single peaked with asymmetric bases, extending to beyond
$\pm\!1000$\,\kms. We measured the mean central wavelengths of the
broad components and the positions of the line peaks of \hbeta,
\hgamma, and \heii. The former has a velocity amplitude of
$520\pm24$\,\kms\ and reaches maximum positive radial velocity at
$\phi\!=\!0.91\pm0.02$. This phasing is consistent with the plasma
motion in the magnetically guided stream that leads to pole\,1 and
away from the observer at an azimuth of $\psi\sim35^\circ$, measured
from the line connecting the two stars.  The line peak displays a
small radial velocity with zero crossing near $\phi\!=\!0.80$. This
component could arise from the ballistic accretion stream near L$_1$.

\begin{figure}[t]
\includegraphics[height=89.0mm,angle=270,clip]{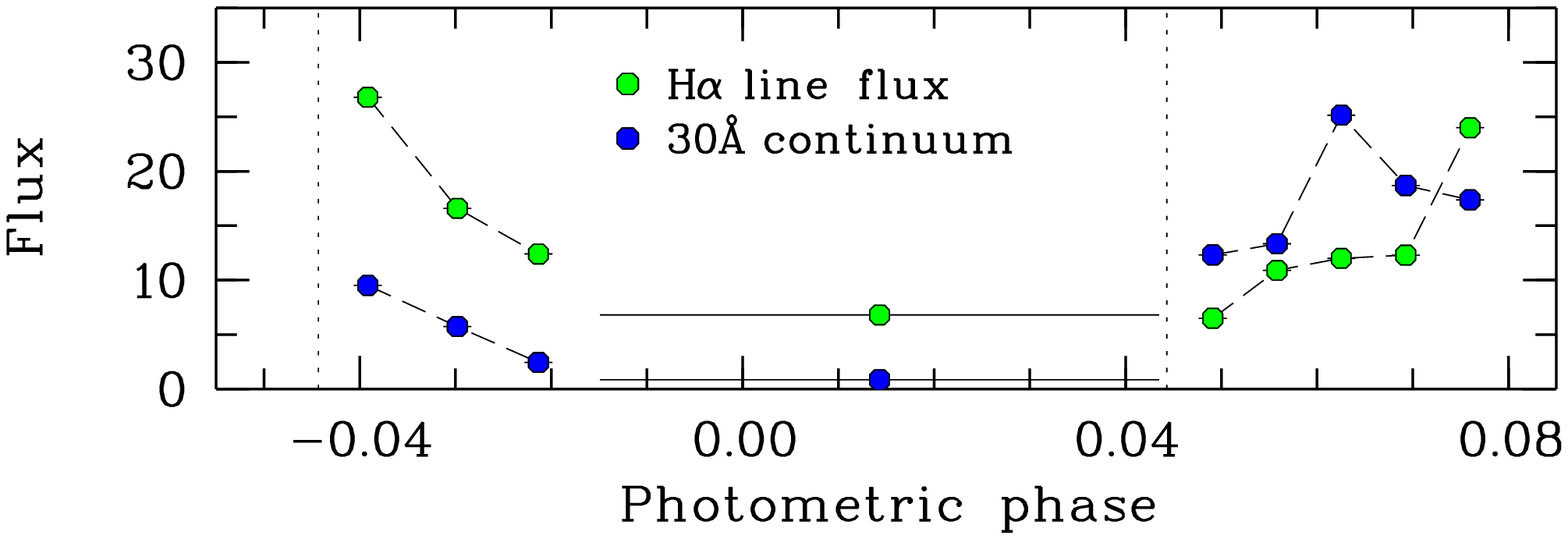}

\vspace{1.0mm}
\includegraphics[height=89.0mm,angle=270,clip]{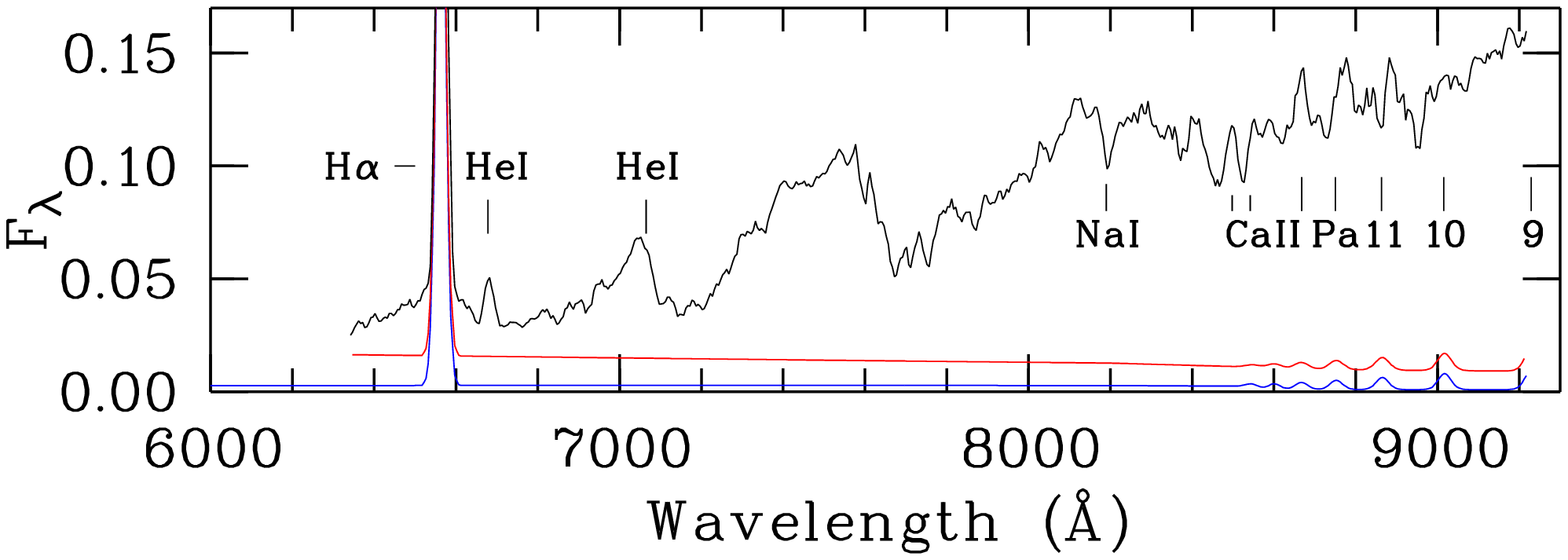}

\vspace{0.0mm}
\includegraphics[height=89.0mm,angle=270,clip]{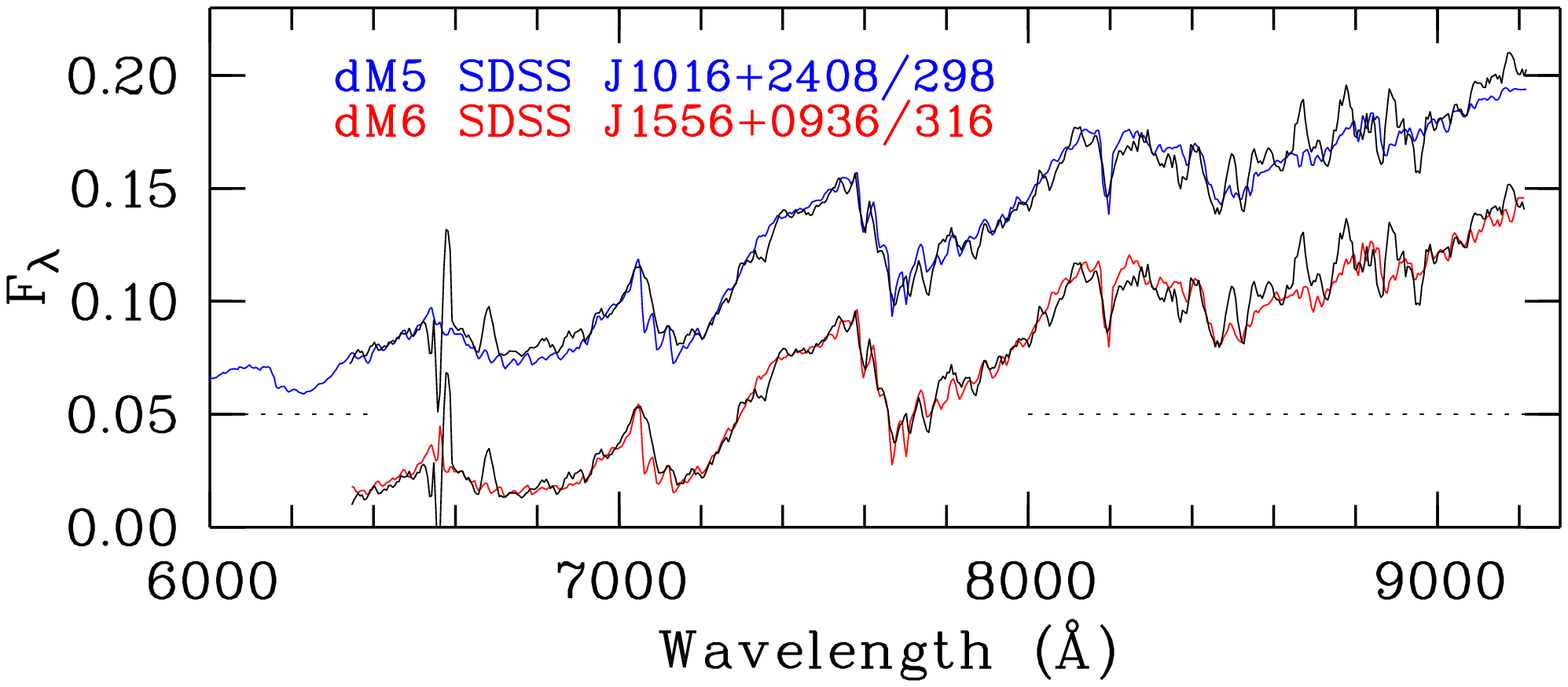}
\caption[chart] {\emph{Top: } Eclipse light curve of the \halpha\
  emission line flux in \ergs\ and of the underlying continuum flux
  observed on 20 November 2000. \emph{Center: } Eclipse spectrum
  (black curve) thermal hydrogen spectrum adjusted to fit the \halpha\
  line flux (blue or red curve, see text). The ordinate is in units of
  $10^{-16}$~ergs\,cm$^{-2}$s$^{-1}$\AA$^{-1}$.  \emph{Bottom: }
  Difference spectra on the same color coding fitted by a dM6 star
  (red) and a dM5 star (blue, shifted upward by 0.05 units).}
\label{fig:eclspec}

\vspace{-2mm}
\end{figure}

\subsection{Spectrum of the secondary star in eclipse}
\label{sec:vlt}

On 20 November 2000, we acquired a spectrum of the secondary star
during the WD eclipse, using the ESO VLT UT1 with FORS1 and grism
G300I (Table~\ref{tab:spec}). The 600\,s exposure was preceded by
three and followed by five 30-s exposures. The run started shortly
after the ingress of the accretion spot into eclipse and extended
until after its egress. The dotted vertical lines in
Fig.~\ref{fig:eclspec}, top panel) indicate the duration of the
eclipse. The 600~s exposure covered the phase interval
$\phi\!=\!-0.0149$ to 0.0435, beginning after the stream component
subsided and ending just before the spot starts to reappear at
$\phi\!=\!0.0438$. The figure shows the light curves of the \halpha\
line flux and of the continuum near \halpha\ integrated over
30\,\AA. The \halpha\ emission stays finite in the eclipse.

The center panel of Fig.~\ref{fig:eclspec} shows the eclipse spectrum
dereddened with the galactic extinction $A_\mathrm{V}\!=\!0.172$
\citep{schlegeletal98}, where we have assumed that \hyeri\ is located
outside the principal dust layer of the galactic disk
\citep{jonesetal11}. The secondary has a dereddened AB magnitude
$i\!=\!20.99$ with an estimated systematic error of 0.10 mag from the
absolute flux calibration of the spectrophotometry.  The spectrum
shows the TiO features characteristic of a late dM star and a strong
\halpha\ emission line with a line flux of
$\ten{6.8}{-16}$\,\ergs. The line is centered approximately at the
rest wavelength and has a velocity dispersion of $\sim\!1000$\,\kms.
It may arise from a tenuous uneclipsed section of the accretion
stream. Regardless of its origin, the line emission will be
accompanied by a thermal continuum that we need to define and subtract
before a spectral type can be assigned to the secondary star.  To this
end, we calculated spectra of an isothermal hydrogen plasma of finite
optical depth, added line broadening, and normalized the spectra to
fit the observed \halpha\ line flux. The free parameters of the model
are the electron temperature $T_\mathrm{e}$, the pressure $P$, and the
geometrical thickness $x$ of the emitting plasma. In the center panel
of Fig.~\ref{fig:eclspec}, we show two examples that bracket the
permitted range of the flux of the sought-after continuum. The blue
spectrum for $T_\mathrm{e}\!=\!10000$\,K, $P\!=\!10$\,dyne\,cm$^{-2}$,
$x\!=\!10^8$\,cm features a low thermal continuum and the red curve
for $T_\mathrm{e}\!=\!20000$\,K, $P\!=\!200$\,dyne\,cm$^{-2}$,
$x\!=\!10^8$\,cm a high one. The bottom panel of
Fig.~\ref{fig:eclspec} shows the observed spectrum with either one of
the model spectra subtracted. Employing a set of dereddened spectra of
SDSS dM stars, we find that the observed spectrum corrected with the
low thermal continuum is best fitted with the dM5 star
SDSS\,J101639.10+240814.2 adjusted by a factor of 287 (rms spectral
flux deviation 0.0055 in the ordinate unit of Fig.~\ref{fig:eclspec},
bottom panel). The corresponding spectrum for the high thermal
continuum is equally well fitted with the dM6 star
SDSS\,J155653.99+093656.5 adjusted by a factor of 289 (rms deviation
0.0053). The two cases tap the full range of thermal continua
permitted by the eclipse spectrum and spectral types outside the
dM5--6 slot quickly fail to provide an acceptable fit. For the dM5
case, the dereddened spectrum of the secondary corrected for the small
thermal contribution has $i\!=\!21.03$ and colors $r\!-\!i\!=\!1.73$
and $i\!-\!z\!=\!0.95$. For the dM6 case with its larger thermal
component, we find $i\!=\!21.22$ with $r\!-\!i\!=\!2.01$ and
$i\!-\!z\!=\!1.10$~\footnote{We folded the SDSS spectrum over the
  Sloan $i$ filter curve for airmass zero, obtaining $i\!=\!14.89$ for
  SDSS\,J101639+240814 and $i\!=\!15.07$ for
  SDSS\,J155653+093656. Adding the distance moduli yields the quoted
  magnitudes. The $i$-band magnitudes of the SDSS spectra represent
  the appropriate reference for the present purpose, although they
  differ by +0.05 and $-0.05$, respectively, from the DR15
  photometry. The quoted colors $r\!-\!i$ and $i\!-\!z$ are those of
  the SDSS photometry.}. The angular radius $R_2/d$ of the secondary star and
its brightness are related by the surface brightness
\begin{equation}
  S = m+5\,\mathrm{log}\,\left[(R_2/d)(10\,\mathrm{pc}/R_\odot)\right],
\label{eq:surf}
\end{equation}
where $m$ is the magnitude of the star in the selected band, $R_2$ its
radius, and $d$ its distance. We calibrated the surface brightness
$S_\mathrm{i}$ in the $i$-band as a function of color, using the
extensive data set of \citet{mannetal15} that combines Sloan
\emph{griz} photometry and stellar parameters. The desired relations
are
\begin{eqnarray}
S_\mathrm{i} & = & 5.34+1.554\,(r-i) \qquad \mathrm{with} \qquad \sigma(S_\mathrm{i})\!=\!0.086 \\
S_\mathrm{i} & = & 5.33+2.829\,(i-z) \qquad \mathrm{with} \qquad \sigma(S_\mathrm{i})\!=\!0.072,
\end{eqnarray}
valid for $r\,-\,i\!=\!1.0-2.6$ and $i\,-\,z\!=\!0.6\!-\!1.3$,
respectively. The quoted standard deviations describe the average
spread of $S_\mathrm{i}$ around the fit. With the colors of the
secondary star quoted above, we obtained mean values from Eqs.~(4) and
(5) of $S_\mathrm{i}(\mathrm{dM5})\!=\!8.02$ and
$S_\mathrm{i}(\mathrm{dM6})\!=\!8.44$. From Eq.~(3), the distance $d$
in pc is given by
\begin{equation}
  \mathrm{log}\,d=(i-S_\mathrm{i})/5+1+\mathrm{log}(f_\mathrm{back}R_2/R_\odot)
\label{eq:dist}
\end{equation}
where $i$ is the magnitude of the continuum-corrected eclipse
spectrum, $S_\mathrm{i}$ the respective surface brightness, $R_2$ the
volume equivalent radius of the Roche lobe, and
$f_\mathrm{back}\!\simeq\!0.961$ the reduction factor for the backside
view of the lobe\footnote{The cross section of the secondary as seen
  along the line connecting the two stars is taken as elliptical with
  axes $y_4$ and $z_6$, yielding
  $f_\mathrm{back}\!=\!\sqrt(y_4\,z_6)/(r^*)_2\!\simeq\!0.961$, in
  Kopal's (1959) notation, with $(r^*)_2$ the equivalent volume
  filling radius of the Roche lobe.}. The dM5--dM6 differences in
$S_\mathrm{i}$ and in the $i$-band magnitude compensate in part,
leading to similar distances with a ratio
$d_\mathrm{dM5}/d_\mathrm{dM6}\!=\!1.11$. We employ Eq.~\ref{eq:dist}
in Sect.~\ref{sec:system}, using the radii of the secondary star
derived from our dynamic models of \hyeri. The error in the mean
dM5--dM6 distance includes the $\pm\,0.10$ mag uncertainty in the flux
calibration, the $\pm\,0.08$ mag scatter in $S_\mathrm{i}$, and half
the difference of the dM5 and dM6 distances. Added quadratically, the
error in $d$ is $\pm\,8.1$\% plus any error that arises from $R_2$.

We estimated the effective temperature of the secondary star from the color
dependence of \teff\ in the data of \citet{mannetal15}, which yields
\teff$\,\simeq\!3070$\,K and 2900\,K for the dM5 and the dM6 case,
respectively. With the best-fit system parameters of
Sect.~\ref{sec:system}, the luminosity of the secondary star becomes
$\ten{2.6\!-\!2.0}{31}$\,\erg\ for a spectral type range of
dM$5\!-\!6$.

\section{Circular spectropolarimetry}
\label{sec:circ}

\begin{figure*}[t]
\includegraphics[bb=181 505 448 716,height=59.5mm,angle=0,clip]{36626f5a.ps}
\hspace{-3mm}
\includegraphics[bb=168 505 383 716,height=59.5mm,angle=0,clip]{36626f5b.ps}
\hfill
\raisebox{59.5mm}{
\includegraphics[width=59.5mm,angle=270,clip]{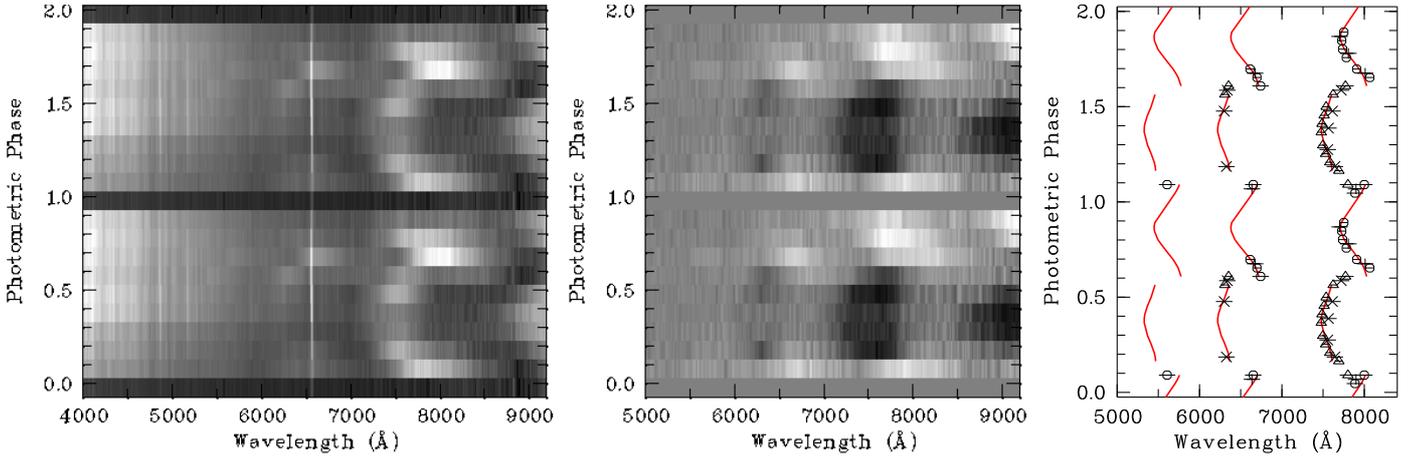} }
\caption[chart]{\emph{Left: } Intensity spectra of the 2008
  spectropolarimetry with the contribution of the secondary star
  subtracted. The mean spectra of two orbits are shown twice for
  visual continuity. Two systems of cyclotron lines are visible that
  originate near two poles of similar field strength. \emph{Center}:
  Circular polarization spectra with white indicating positive and
  black negative polarization.  \emph{Right:} Maxima of the cyclotron
  lines of both poles (circles, triangles), extrema of the circular
  polarization (plus signs, crosses), and best-fit line positions as
  functions of orbital phase (red curves).}
\label{fig:gray}
\end{figure*}

\begin{figure*}[t]
\includegraphics[height=89.0mm,angle=270,clip]{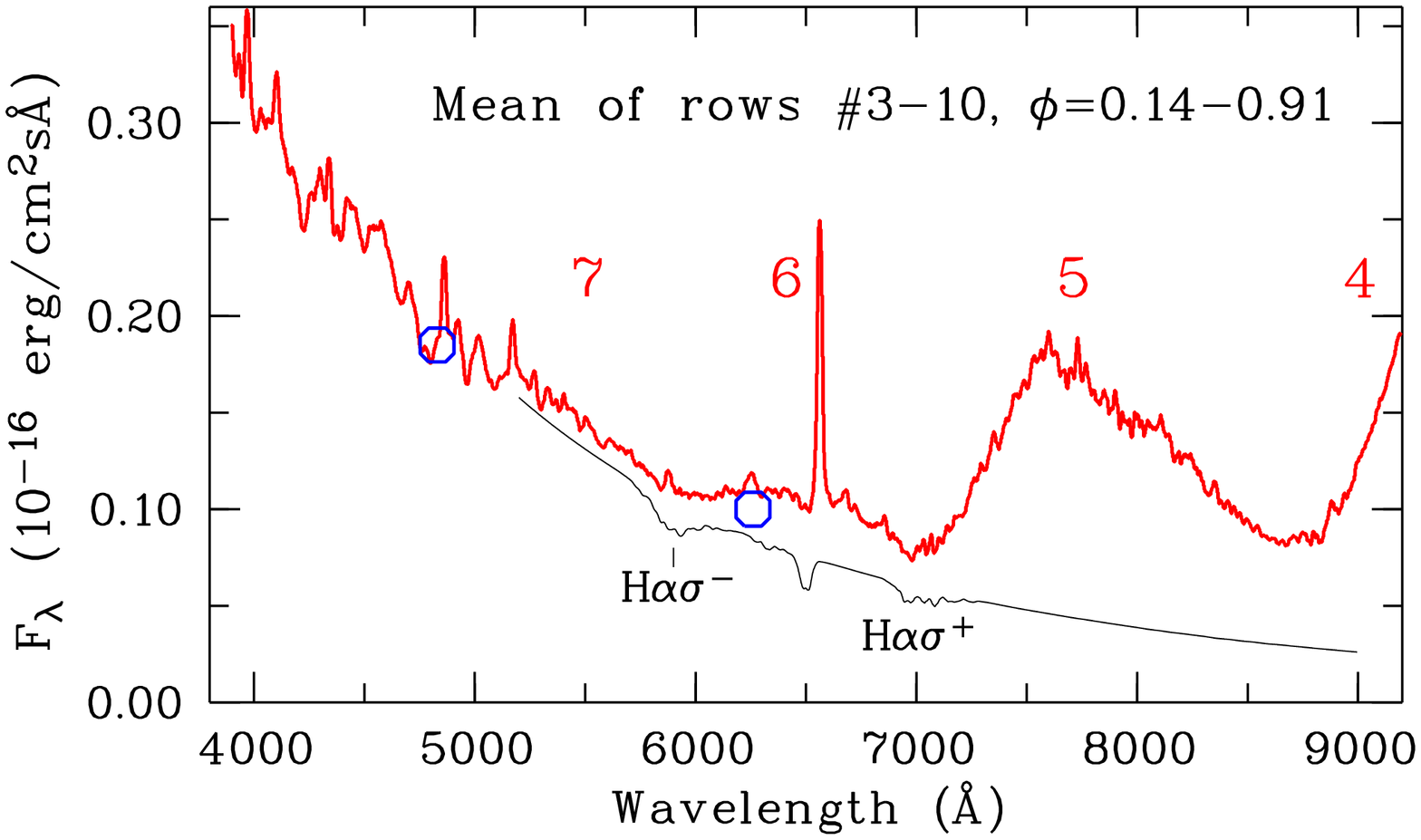}
\hfill
\includegraphics[height=90.0mm,angle=270,clip]{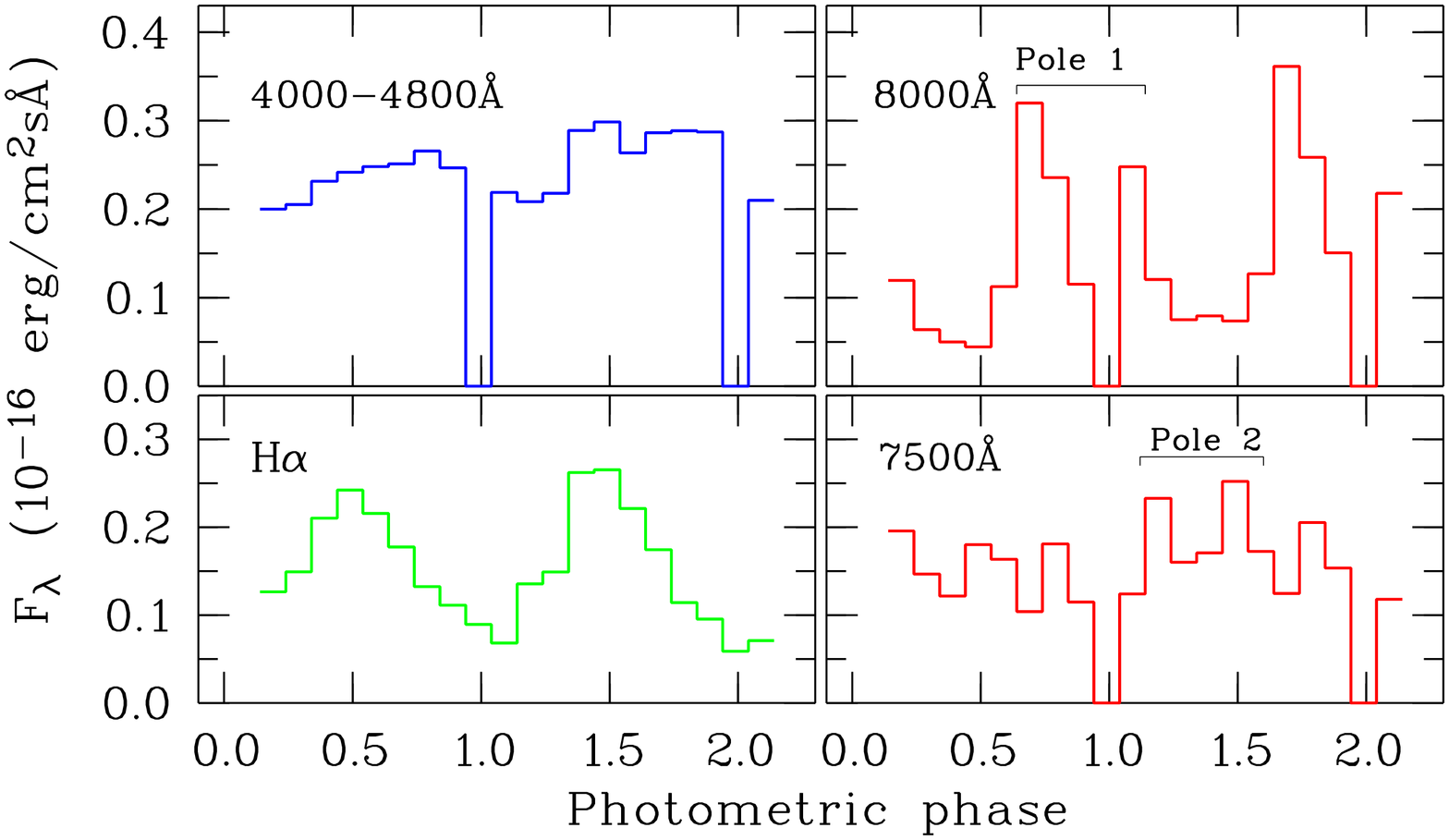}
\caption[chart] {\emph{Left:} Mean bright-phase dereddened intensity
  spectrum of the 2008 spectropolarimetry with the contribution of the
  secondary star subtracted (red curve). For comparison, the
  appropriate model spectrum of a magnetic WD is shown (black
  curve). The mean fluxes of the November 2017 and February 2018 $gr$
  photometry are included as the two open blue circles. The red
  numbers indicate the cyclotron harmonics.  \emph{Right:} Light
  curves for selected wavelength intervals derived from spectral
  set\,1 before phasefolding, but after subtraction of the secondary
  star. The individual panels show the quasi-B band flux, the
  cyclotron beaming of the pole-1 and pole-2 emissions, and the
  \halpha\ line flux (in arbitrary units). }
\label{fig:meanspec}

\vspace{-1mm}
\end{figure*}

\begin{figure*}[t]
\includegraphics[height=86.0mm,angle=270,clip]{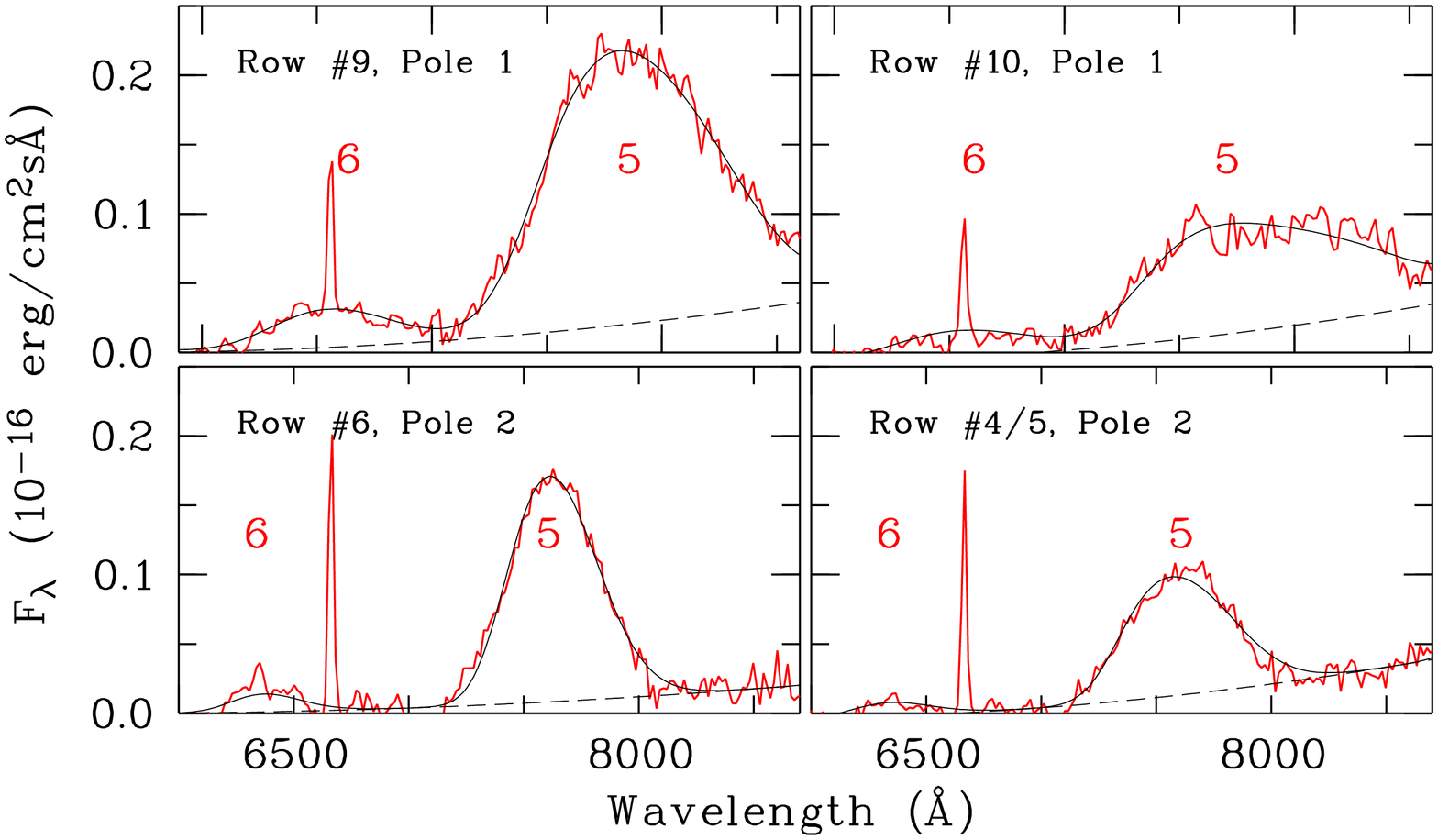}
\hfill
\includegraphics[height=42.0mm,angle=270,clip]{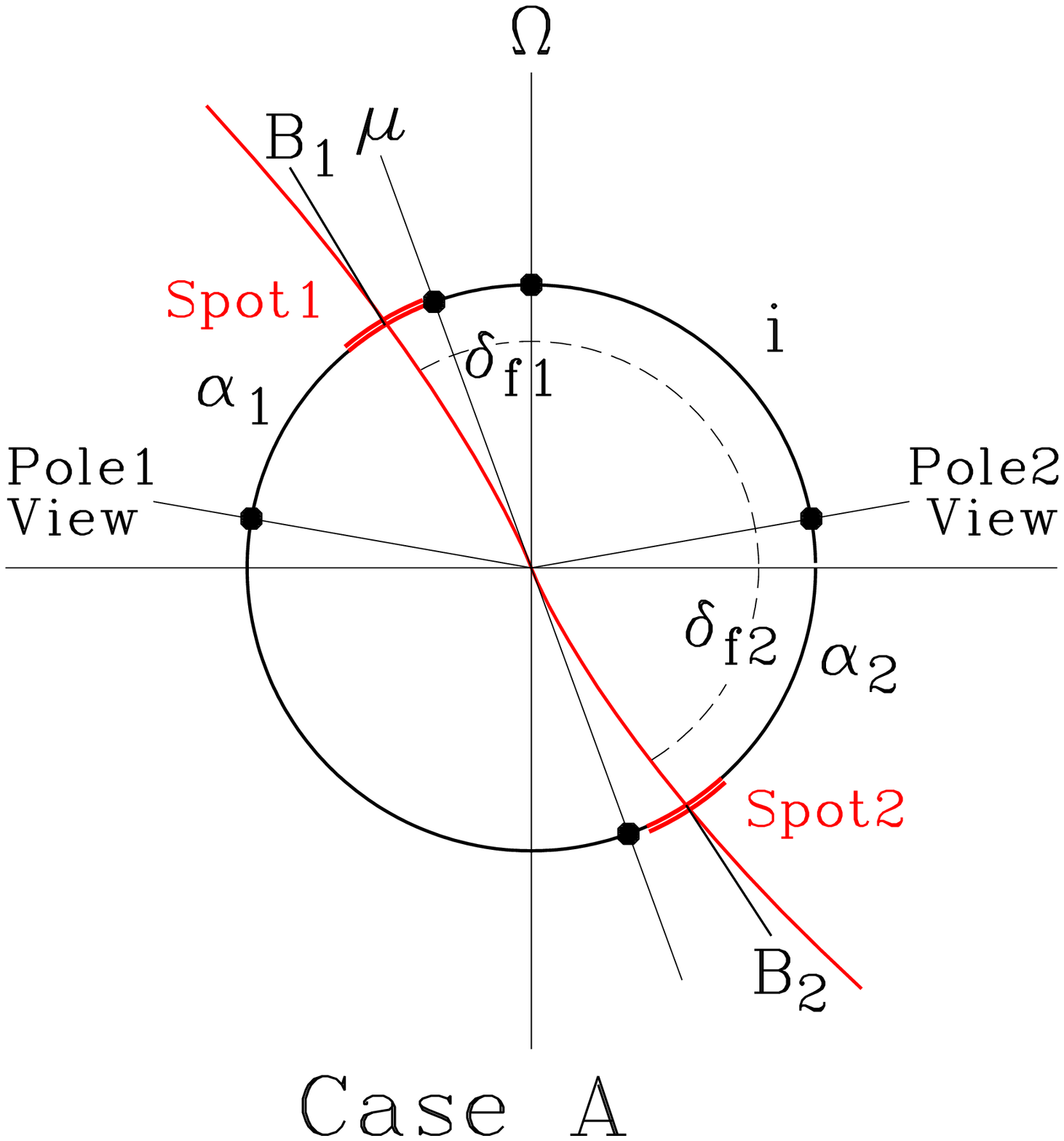}
\hspace{3mm}
\includegraphics[height=42.0mm,angle=270,clip]{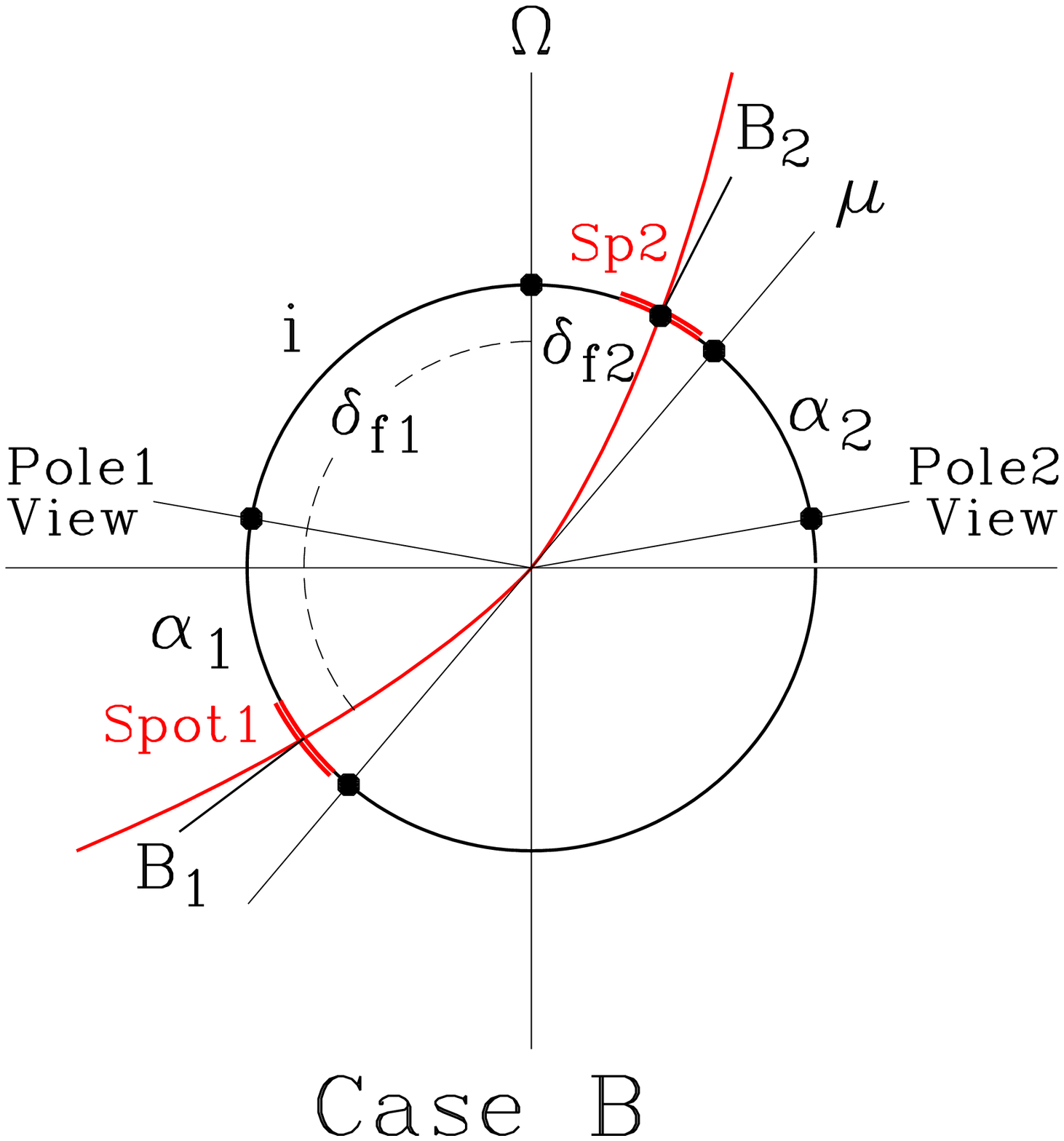}
\caption[chart] {\emph{Left:} Cyclotron spectra of poles 1 and 2 (red
  curves) least-squares fitted with constant-temperature models for
  the plasma parameters listed in Table~\ref{tab:cyc} (black
  curves). \emph{Right panels:} Magnetic geometry of the WD for case A
  and case B accretion (see text). The secondary star is located to
  the left. Angles are measured between the dots on the circumference.
  Dipolar field lines are included as red curves.}
\label{fig:cyc}

\end{figure*}

We studied the magnetic field of the WD in \hyeri\ by phase-resolved
circular spectropolarimetry performed on 31 December 2008, when the
system was accreting at a low level. A total of 40 sets of ordinary
and extraordinary spectra were obtained with the ESO VLT UT2 equipped
with FORS1 and GrismG300V, covering two consecutive orbital periods
(Table~\ref{tab:spec}). The pipeline reduction combines two spectra
each for two positions of the $\lambda/4$ plate in order to account
for possible cross-talk from linear polarization. This procedure
yielded set\,1 of 20 calibrated intensity and circular polarization
spectra, each covering a phase interval $\Delta
\phi\!=\!0.10$. Spectral features faithfully repeated in both orbits
and we phasefolded the intensity and polarization spectra for further
analysis. In case of the intensity spectra, the phase resolution can
be improved to $\Delta \phi\!=\!0.05$ by adding the individual
ordinary and extraordinary spectra to form set\,2 of 40 provisionally
calibrated intensity spectra. Set\,1 was employed for a quantitative
analysis of the spectral flux, set\,2 for tracing the motion of the
cyclotron line peaks and measuring the \halpha\ radial
velocities. Both were corrected for the secondary star, using the dM5
representation of the eclipse spectrum of Fig.~\ref{fig:eclspec}
extrapolated into the blue spectral region and set\,1 was dereddened
with $A_\mathrm{V}\!=\!0.172$ \citep{schlegeletal98}. \halpha\
emission is much weaker than in the 2000 eclipse spectroscopy of
Sect.~\ref{sec:vlt} and we do not correct for the probably tiny
contribution of the associated thermal continuum.  Grayscale
representations of the corrected phasefolded \mbox{set-1} intensity
and circular polarization spectra are shown in Fig.~\ref{fig:gray},
repeated twice for visual continuity. Rows \#1, \#11, and \#21
represent the eclipse. The orbital mean spectrum outside eclipse (rows
\#3-10, $\phi\!=\!0.14-0.91$) is shown in the left panel of
Fig.~\ref{fig:meanspec} (red curve); an appropriate model spectrum of
the magnetic WD is added for comparison (black curve). The two blue
circles are the mean out-of-eclipse $g$ and $r$-band fluxes of our
2017 and 2018 photometry, corrected for the secondary. They show that
the 2008, 2017, and 2018 observations were performed in similar states
of low
% accretion. The right panels of Fig.~\ref{fig:meanspec} show blue and
% red light curves extracted from the set\,1 spectra. The red ones are
accretion. The right panels of Fig.~\ref{fig:meanspec} show the light
curves for the blue and red wavelength intervals extracted from the
set\,1 spectra. The red ones ($7400-7600$\,\AA\ and $7900-8100$\,\AA)
are shaped by the beamed optically thin cyclotron emission of poles 1
and 2. The blue one ($4000-4800$\,\AA) represents the sum of the
photospheric and spot emissions of the WD.~Disentangling the two
components 
%to the blue light curve 
proved infeasible. The
\halpha\ line emission is discussed in Sect.~\ref{sec:nelhy}.

\subsection{Cyclotron spectroscopy}
\label{sec:cyc}

The cyclotron lines of pole\,1 are visible from $\phi\!=\!0.6\!-\!1.1$
with positive circular polarization and those of pole\,2 in the
remainder of the orbit with negative circular polarization. Overlaps
between the two line systems occur in rows \#3 ($\phi\!=\!0.14-0.23$)
and and \#7 ($\phi\!=\!0.54\!-\!0.63$), where the spectra show the
signatures of both poles. We extracted the cyclotron lines from the
individual intensity spectra by removing the underlying WD
continuum. The cyclotron line profiles were then subjected to
least squares fits using the theory of \citet{chanmugamdulk81} for an
isothermal plasma.  The free parameters of the model are the plasma
temperature k$T$, the field strength $B$, the viewing angle $\theta$
against the field direction, the thickness parameter $\Lambda$\, and a
remnant optically thick continuum represented by a second-order
polynomial (dashed lines in Fig.~\ref{fig:cyc}). The thickness
parameter $\Lambda$ of cyclotron theory is related to the column
density $x_\mathrm{s}$ of the cooling region by $\Lambda\!=\!4\pi
ex_\mathrm{s}/\mu_\mathrm{e}m_\mathrm{u}B$, where $e$ is the
elementary charge, $\mu_\mathrm{e}$ the number of electrons per
nucleon in the plasma, $m_\mathrm{u}$ is the atomic mass unit, and $B$
the field strength.
 
\begin{table}[b]
\begin{flushleft}
  \caption{Physical parameters of spots 1 and 2 derived from
    least squares fits of the model of an isothermal plasma to the
    cyclotron line profiles.}

\begin{tabular}{c@{\hspace{2.0mm}}c@{\hspace{2.0mm}}c@{\hspace{2.0mm}}c@{\hspace{2.0mm}}c@{\hspace{2.2mm}}c@{\hspace{2.2mm}}c@{\hspace{2.2mm}}c@{\hspace{2.2mm}}c}\\[-2ex]
\hline\hline \\[-1.5ex]
Pole & Row  & $\phi $ & Harmonics& $B_{sp}$ & k$T$   & $\langle \theta\rangle$ & log\,$\Lambda$ & $\dot m$  \\
&  \#  & &  fitted  &    (MG)    &  (keV) &   (\degr)  &  & (g/cm$^{2}$s)               \\[1.0ex]
\hline\\[-1ex]
1   &  2  & 0.07 & (4) 5 6  &   27.50    &  2.25  &    63  &   5.84 & 0.0095 \\
1   &\hspace{-1.7mm}10& 0.87 & (4) 5 6  &   27.37    &  2.20  &    49  &   5.81 & 0.0090  \\
1   &  9  & 0.78 & (4) 5 6  &   27.51    &  2.25  &    59  &   5.84  & 0.0102\\[1.0ex]
2   &  6  & 0.48 & (4) 5 6  &   28.83    &  1.70  &    68  &   5.55 &0.0048 \\
2   &\hspace{-1.2mm}4\,/\,5 & 0.33 & (4) 5 6  &   28.88    &  1.65  &    62  &   5.50 & 0.0040 \\[1.0ex]
\hline\\
\end{tabular}
\label{tab:cyc}
\end{flushleft}

\vspace{-3mm}
\end{table}

Because of trade-off effects between k$T$ and log\,$\Lambda$, reliable
values of these two parameters can not be determined without
additional information. We took recourse to the results of the
two-fluid radiation-hydrodynamic cooling theory of
\citet{fischerbeuermann01} and explain our approach for the case of
the \mbox{pole-2} spectrum of row \#6 in the lower left panel of
Fig.~\ref{fig:cyc}. Good fits to that spectrum were obtained along a
narrow valley in the $\Lambda\!-\!\mathrm{k}T$ plane that follows
log\,$\Lambda\!=\!6.64\!-\!4.73\,$\,log(k$T$) and extends to
parameters quite inappropriate for a post-shock region dominated by
cyclotron cooling. Cooling theory provides a second relation between
log$\Lambda$ and log(k$T$) that runs nearly orthogonal to that of the
line fits. For the case of the \#6 spectrum, it reads
log\,$\Lambda\!=\!4.97\!+\!2.50$\,log(k$T$)\,\footnote{The second
  $\Lambda$-k$T$ relation is based on equations for the post-shock
  plasma temperature k$T$ and the column density $x_\mathrm{s}$ of the
  post-shock cooling flow presented in Figs.\, 5 and 6 and Eqs.\,(19)
  and (20) of \citet{fischerbeuermann01}. Elimination of the variable
  $\dot m\,B_7^{-2.6}$, with $\dot m$ the mass-flow density in
  g\,cm$^{-2}$s$^{-1}$ and $B_7$ the field strength in units of
  $10^7$\,G, yields the desired relation between k$T$ and
  $\Lambda(x_\mathrm{s})$, which is well fitted by a power law valid
  for k$T$ up to about 3\,keV. For the present purpose, the relations
  in question were re-calculated for a plasma of solar composition and
  a WD mass $M_1\!=\!0.4$\,\msun\ as suggested by our dynamic models in
  Sect.~\ref{sec:system}.}. 
The intersection of the two relations
defines the most probable values of log$\Lambda$ and k$T$, yielding
k$T\!=\!1.70$\,keV and log\,$\Lambda\!=\!5.55$. Table~\ref{tab:cyc}
lists the corresponding fits to all spectra that can uniquely be
assigned to either pole\,1 or pole\,2. The emerging picture is that of
two emission regions with similar field strengths
$B_\mathrm{sp}\!\simeq\!27.5$\,MG and 28.8\,MG, thickness parameters
log\,$\Lambda\!\simeq\!5.8$ and 5.5, and temperatures of
k$T\!\simeq\!2.2$\,keV and 1.7\,keV for poles 1 and 2, respectively.
The third parameter derived from the cyclotron fits is the mean
viewing angle $\langle \theta\rangle$ between the line
of sight and the direction of the accreting field line averaged over
the spot.  Closest approach to the field line occurs for pole\,1 at
$\phi\!\simeq\!0.87$ in row \#10 and for pole\,2 at
$\phi\!\simeq\!0.33$ in rows \#4 and \#5 with minimum viewing angles
of $\langle \theta_\mathrm{min,1} \rangle\!\simeq\!49\degr$
and $\langle \theta_\mathrm{min,2}
\rangle\!\simeq\!62\degr$ for spots\,1 and 2,
respectively. Since the cyclotron lines widen and weaken rapidly with
decreasing $\theta$, the quoted angles may somewhat overestimate the
true mean values.

Table~\ref{tab:cyc} also lists the mass flow densities $\dot m$ that
are delivered also by two-fluid cooling theory.  They fall far below
the $\sim\!1$\,g\,cm$^{-2}$s$^{-1}$ of a bremsstrahlung-dominated
emission region and are only about an order of magnitude away from the
transition to the non-hydrodynamic regime of the bombardment solution
\citep{woelkbeuermann92,fischerbeuermann01}.

Complementary information on $\theta$ is obtained from the orbital
motion of the cyclotron line peaks and of the circular polarization
extrema. The peak wavelengths measured from the spectra of sets\,2 and
1, respectively, are shown in the right panel of
Fig.~\ref{fig:gray}. Near $\phi\!=\!0.1$ and 0.6, both line systems
overlap and are difficult to disentangle. As in Paper~I, the motion of
the cyclotron lines was modeled, using a parameterized form of the
frequencies of optically thin harmonics in units of the cyclotron
frequency as functions of k$T$ and $\theta$. We defined the field
vectors in the two spots as $\vec{B_\mathrm{sp,1}}$ and
$\vec{B_\mathrm{sp,2}}$ and obtained field strengths and directions by
least-squares fitting the phase-dependent motion of the line peaks of
the fifth and sixth harmonics (red curves in Fig.~\ref{fig:gray}). As
input we used the plasma temperatures of Table~\ref{tab:cyc} and an
inclination of $i\!=\!80\degr$ from Table~\ref{tab:system}. The
results are presented in Table~\ref{tab:harm}, where we list the field
strengths $B$, the azimuth angles $\psi_\mathrm{f}$, and the
colatitudes $\delta_\mathrm{f}$ of the accreting field lines. The
results are quoted for two accretion geometries with pole\,1 either in
the ``southern'' hemisphere below the orbital plane (1S) or in the
``northern'' one above it (1N). Note that the fit does not provide
information on the location of the spots and the orientation of the
magnetic axis. The angle between the field vectors
$\vec{B_\mathrm{sp,1}}$ and $\vec{B_\mathrm{sp,1}}$ is not far from
180\degr, at least in the 1N--2S case. Combined with the fact, that
both spots display circular polarization of opposite sign, the data
suggest a field structure that is dominated by a dipole and possibly
octupole rather than a quadrupole.

\begin{table}[b]
\begin{flushleft}
  \caption{Geometry of the accreting field lines derived from
    least squares fits to the orbital motion of the cyclotron
    lines (Fig.~\ref{fig:gray}). }

\begin{tabular}{@{\hspace{3.0mm}}c@{\hspace{6.0mm}}c@{\hspace{6.0mm}}c@{\hspace{6.0mm}}c@{\hspace{6mm}}c@{\hspace{3mm}}c@{\hspace{3mm}}cc}\\[-2ex]
\hline\hline \\[-1.5ex]
Pole   & $B_\mathrm{sp}$  & $\delta_\mathrm{f}$ & $\psi_\mathrm{f}$ & $\theta_{min}$ & $\angle (\vec{B_\mathrm{sp,1}},\vec{B_\mathrm{sp,2}})$&$\Delta(\phi)$\\
       & (MG) &  (\degr)             &  (\degr)             &  (\degr)   &   (\degr)                       &                               \\[1.0ex]
\hline\\[-1.0ex]
 1\,N  & 27.2 &  36              &               49 &  45    &                               &                               \\ 
 2\,S  & 28.6 &\hspace{-2.2mm}136&\hspace{-2.2mm}224&  56    &\raisebox{1.5ex}[-1.5ex]{170.9}&\raisebox{1.5ex}[-1.5ex]{0.487}\\[1.0ex]
 1\,S  & 27.3 &\hspace{-2.2mm}126&               57 &  46    &                               &                               \\ 
 2\,N  & 28.4 &  28              &\hspace{-2.2mm}227&  52    &\raisebox{1.5ex}[-1.5ex]{153.2}&\raisebox{1.5ex}[-1.5ex]{0.472}\\[1.0ex]
\hline\\                                                
\end{tabular}                                        
\label{tab:harm}                                         
\end{flushleft}                                         
                                                        
\vspace{-3mm}                                           
\end{table}

\begin{figure*}[t]
\includegraphics[width=60.0mm,angle=270,clip]{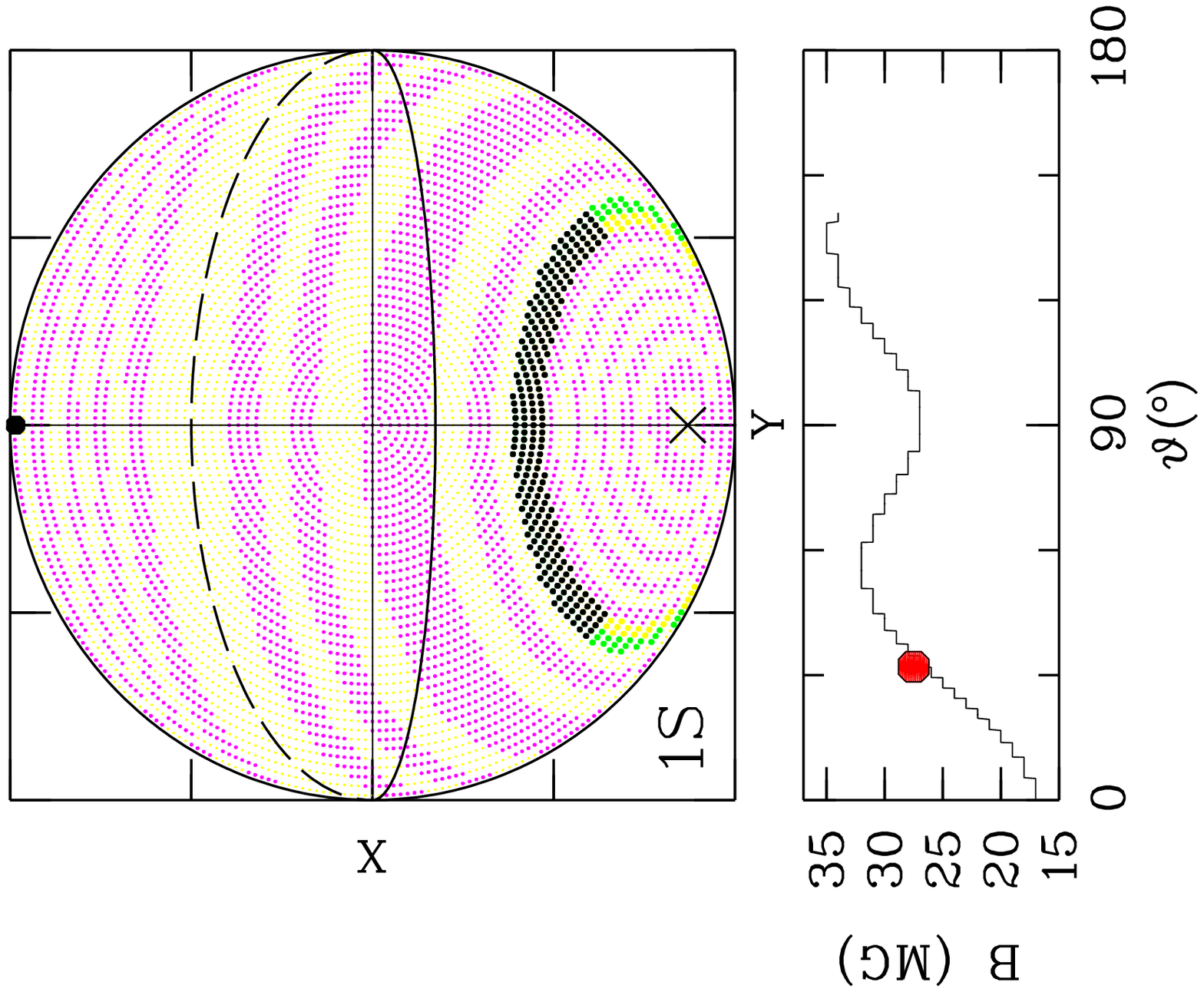}
\hspace{-1.0mm}
\includegraphics[width=60.0mm,angle=270,clip]{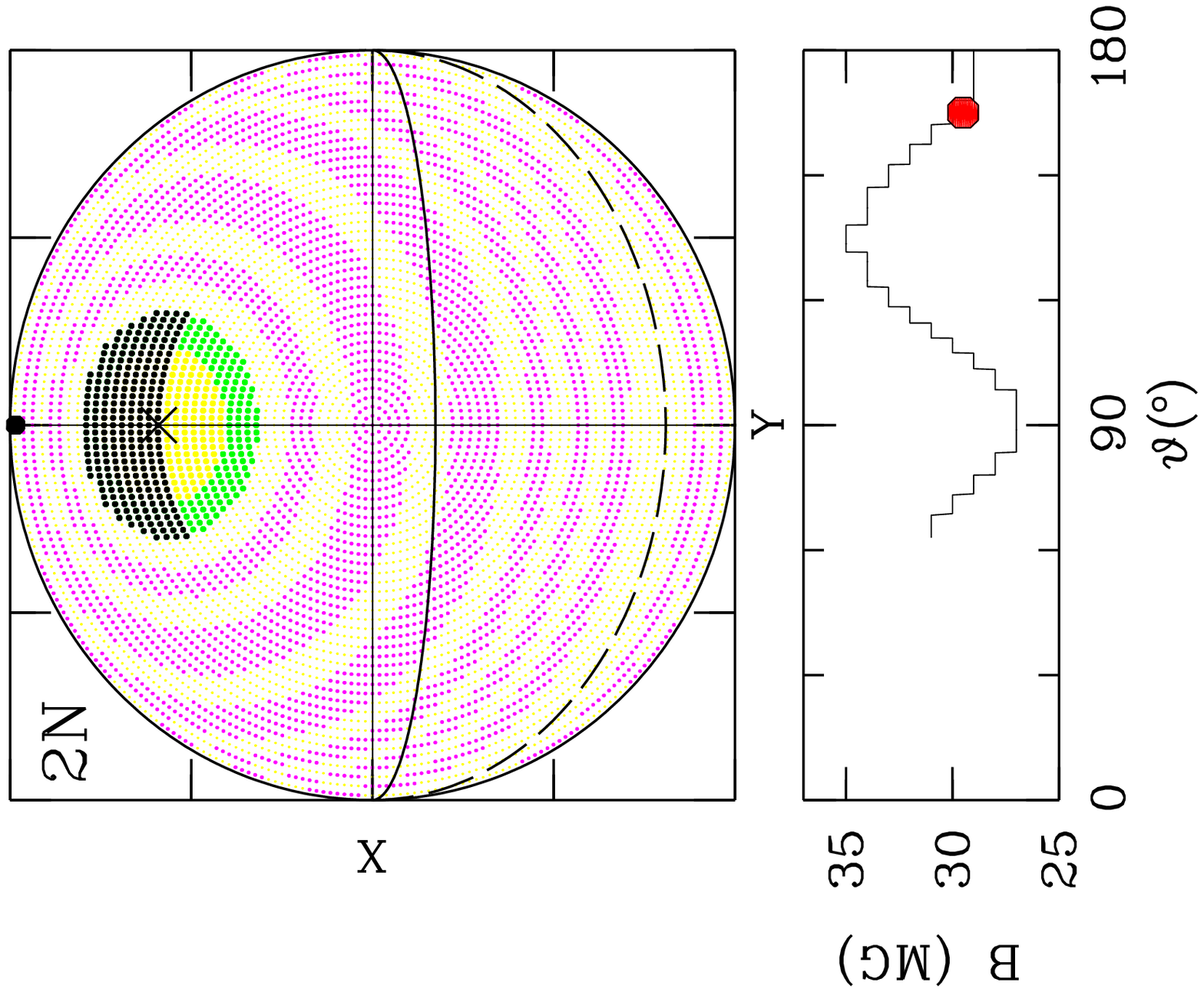}
\hspace{0.0mm}
\includegraphics[height=86.3mm,angle=270,clip]{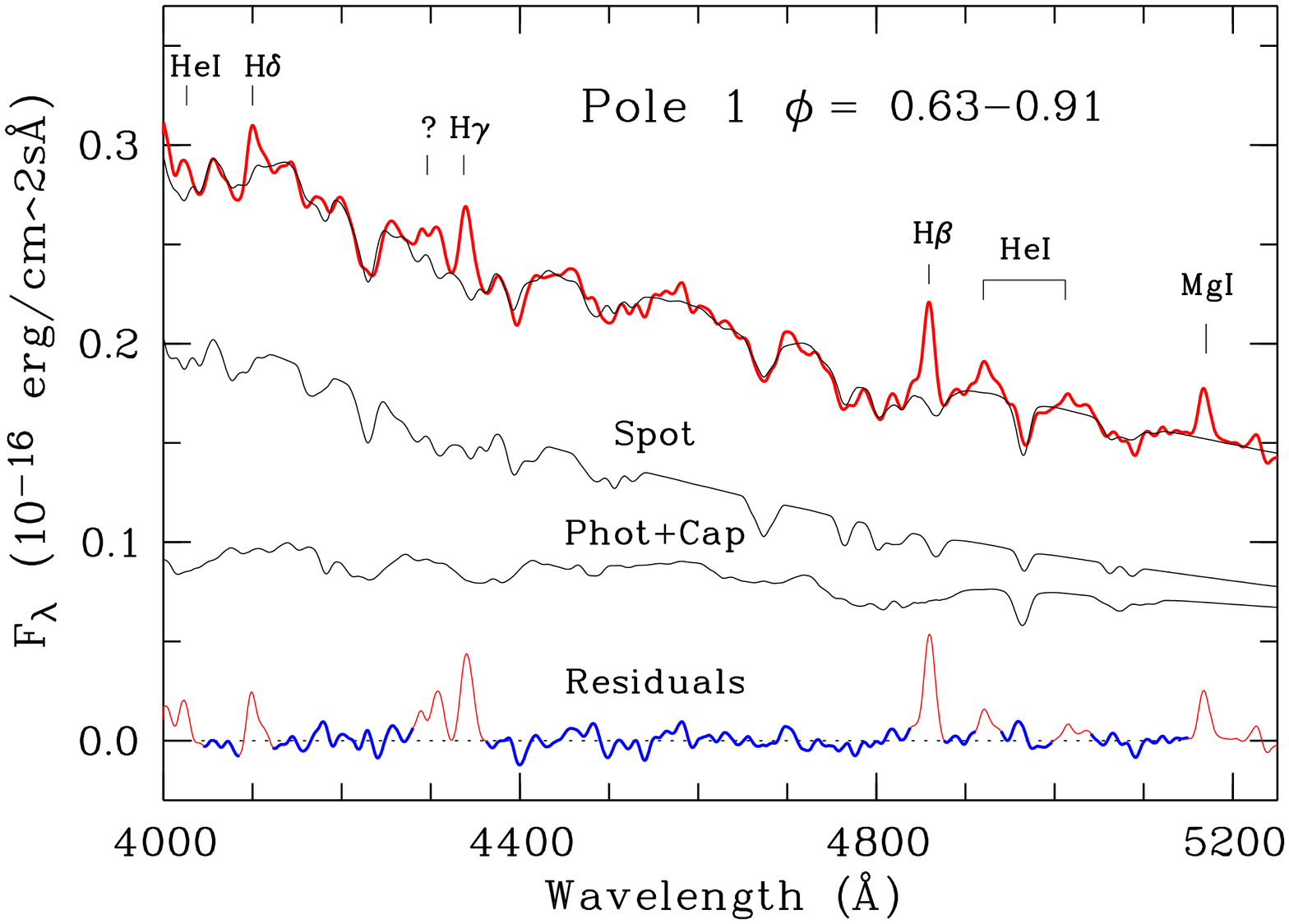}
\caption[chart] {\emph{Left panels: }Magnetic-field distributions of
  the WD for the case~B multipole model of Table~\ref{tab:zee} and the
  phases of best visibility of spots 1S and 2N. The footpoints of the
  rotational and the magnetic axes are indicated by the black dot and
  the cross ($\times $), respectively.  The magnetic equator and the
  orbital plane are marked by the dashed and the solid black line,
  respectively. The individual projected-area elements are marked by
  tiny dots, selecting or suppressing every second integer field
  strength. The spot field is marked in green and the viewing-angle
  selected spot is emphasized in black. The bottom panels show the
  model field strength vs. the magnetic colatitude $\vartheta$. The
  red dots mark the spot fields. \emph{Right panel: } Observed pole-1
  spectrum (red curve) with the best-fit Zeeman spectrum for the
  adopted multipole model (black curves).  The residuals of the fit
  are displayed at the bottom with the wavelength sections that are
  included in the fit highlighted in blue. }
\label{fig:zee}
\end{figure*}

\subsection{Zeeman spectroscopy}
\label{sec:zee}

The blue continuum in the left panel of Figs.~\ref{fig:gray}
represents the photospheric emission of the WD including a strong spot
component. To facilitate the Zeeman analysis, we transformed the set-1
spectra to the rest system of the WD, employing the preferred
dynamical model of Sect.~\ref{sec:system}.  There is little orbital
variation in the Zeeman lines except near the transitions between the
visibility of poles 1 and 2. We adopted the averages of rows \#8--10
($\phi\!=\!0.63-0.91$) or \#3--6 ($\phi\!=\!0.14-0.52$) as
representative of the hemispheres that include pole\,1 or 2,
respectively. The dominant field strengths are 27--28\,MG in pole\,1
and 29--30\,MG in pole\,2. The 4000--5250\AA\ section of the pole-1
spectrum is shown in Fig.~\ref{fig:zee} (red curve). Unfortunately,
the circular polarization spectra, which contain information on the
direction of the magnetic field vector $\vec{B}$, are too noisy to be
of any use, limiting our ability to distinguish between different
magnetic field structures that fit the intensity spectra similarly
well.

Our spectral synthesis program employs an improved set of the model
Zeeman spectra calculated by \citet{jordan92} and previously used by
\citet{euchneretal02,euchneretal05,euchneretal06} and
\citet{beuermannetal07}. The present version includes the Balmer lines
up to \hdelta\ and consists of the log\,$g\!=\!8$ intensity spectra
for 16 effective temperatures from 8 to 100\,kK, integer field
strengths $B$ from 1 to 100\,MG, and 17 viewing angles
$\theta\!=\!0\degr$ to $180$\degr, uniformly distributed in
cos\,$\theta$.  The model spectra were calculated for a Stark
broadening factor $C\!=\!0.1$ \citep{jordan92}. Interpolation in
$T_\mathrm{eff}$ and in $\theta$ is unproblematic, while interpolation
in $B$ is impracticable.  The spectra were smoothed to match the
observed resolution of 10\,\AA\ FWHM. At this resolution, the 1-MG
spacing is just about adequate and misfits stay small.

We considered a magnetic model that includes the zonal multipole
components of degree $\ell\!=\!1\!-\!3$, that is the aligned dipole,
quadrupole, and octupole, inclined by a common angle $\alpha$ against
the line of sight. For the relative field strengths
$r_\mathrm{dip}\!=\!1\!-\!r_\mathrm{oct}$, $r_\mathrm{qua}$, and
$r_\mathrm{oct}$, the combined polar field strength is
$B_\mathrm{p}\!=\!(1\!+\!r_\mathrm{qua})B_0$, with $B_0$ a scaling
factor. We divided the visible disk of the WD into 6561 limb-darkened
projected area elements, and collected them into 17$n$
field-strength and viewing-angle bins (k,l), $k\!=\!1...n$ for the nearest
integer field strength $B_\mathrm{k}$ and $l\!=\!1...17$ for the
nearest cos\,$\theta_\mathrm{l}$ value. The unreddened spectral flux
at the Earth is
\begin{equation}
  f_\mathrm{\lambda}(B_0,r_\mathrm{oct},\alpha) = (R_\mathrm{wd}/d)^2\sum_\mathrm{k=1}^\mathrm{n}\sum_\mathrm{l=1}^{17}\,a_\mathrm{k,l}\,F_\mathrm{\lambda}(B_\mathrm{k},T_\mathrm{k,l},\theta_\mathrm{l}),
\label{eq:tomo}
\end{equation} 
where $a_\mathrm{k,l}$ is the integrated limb-darkened fractional
projected-area of bin (k,l),
$F_\mathrm{\lambda}(B_\mathrm{k},T_\mathrm{k,l},\theta_\mathrm{l})$
the data bank spectrum for that bin at the interpolated temperature
$T_\mathrm{k,l}$, and $(R_\mathrm{wd}/d)^2$ the dilution factor, with
$R_\mathrm{wd}$ the WD radius and $d$ the distance.  The best values
of $B_0$, $r_\mathrm{qua}$, $r_\mathrm{oct}$, and $\alpha$ were
determined in a grid search, the $T_\mathrm{k,l}$ and the dilution
factor by a formal least squares fit at each grid point. Obtaining a
stable fit, requires a severely restricted number of independent
temperatures.  The spot emits about 2/3 of the blue flux from
$\sim\!5$\% of the area, requiring a minimum of two temperatures,
naturally identified as a high spot temperature $T_\mathrm{sp}$ and a
low photospheric temperature $T_\mathrm{ph}$. Some fits benefit from a
minor third component, such as a warm polar cap with $T_\mathrm{cap}$.
We extended the fit over the wavelength interval $4000\!-\!5200$\,\AA,
excluding sections around the Balmer emission lines, some HeI lines,
and an unidentified line complex around 4300\AA. A formal $\chi^2$ was
calculated for 85 resolution elements of 10\,\AA\ width, using
relative flux errors of 1.8\% for pole\,1 and 2.4\% for pole\,2,
measured from the scatter among the set-1 intensity spectra.

\begin{table*}[t]
\begin{flushleft}
\caption{Physical parameters derived from common Zeeman fits to the
    spectra of poles\,1 and 2 for a quasi-dipole and the best
    multipole model. Cols.~(18) and (19) quote the nominal values of
    $T_\mathrm{ph}$ and $R_\mathrm{wd}/d$ from the Zeeman fit with the
    estimated systematic errors. }

\medskip 
\begin{tabular}{@{\hspace{1.0mm}}c@{\hspace{3.0mm}}c@{\hspace{3.0mm}}c@{\hspace{2.0mm}}c@{\hspace{2.0mm}}c@{\hspace{2.0mm}}c@{\hspace{2.0mm}}c@{\hspace{2.0mm}}c@{\hspace{2.0mm}}c@{\hspace{1.0mm}}c@{\hspace{1.0mm}}c@{\hspace{2.0mm}}c@{\hspace{2.0mm}}c@{\hspace{2.0mm}}c@{\hspace{1.0mm}}c@{\hspace{2.0mm}}c@{\hspace{0.0mm}}c@{\hspace{0.0mm}}c@{\hspace{0.0mm}}c@{\hspace{0.0mm}}c@{\hspace{1.0mm}}r}\\[-5ex]
\hline\hline \\[-1.5ex]
(1) & (2) & (3) & (4) & (5) & (6) & (7) & (8) & (9) & (10) & (11)  & (12) & (13) & (14) & (15) & (16) & (17) & (18) & (19) & (20) & (21) \\
Pole&$r_\mathrm{qua}$&$r_\mathrm{oct}$&$B_0$& $\alpha$&$\delta_\mathrm{\mu}$&$\langle \delta_\mathrm{f}\rangle$ &$\langle \theta_\mathrm{f}\rangle$ &$\langle \vartheta_\mathrm{sp}\rangle$&$\langle \beta_\mathrm{sp}\rangle$ &$\langle f_\mathrm{sp}\rangle$&$B_\mathrm{sp}$& $B_\mathrm{pole}$ & $B_\mathrm{dip}$ & $B_\mathrm{qua}$ & $B_\mathrm{oct}$&$B_\mathrm{min}\!-\!B_\mathrm{max}$&$T_\mathrm{phot}$&$R_\mathrm{wd}/d$&$\chi^2$& Note\\
&              &              &(MG) & (\degr)   & (\degr)             &     (\degr)                       &  (\degr)   &  (\degr)                          &  (\degr)                    &            &         (MG)                 &    (MG)         &     (MG)       &    (MG)        &    (MG)       &     (MG)         &    (kK)      & ($10^9$cm/kpc) &     \\[1.0ex]
\hline\\[-1ex]
\multicolumn{17}{l}{\emph{(a)~~Quasi-dipole:}}  \\ [0.5ex]
1\,N &\hspace{-1.6mm}$-0.05$&$ 0.00$& 38.5 & 37  &               113 &\hspace{1.2mm}44 & 52    & 52 & \hspace{-1.2mm}31 &0.054&$27\!-\!28$&$36.6$ & $38.5$ &\hspace{-1.6mm}$-1.9$ & $0.0$ & $19\!-\!37$   & 9.2\,(0.7) & 1.07\,(11)  &88.6 & (1)\\
2\,S &               0.05   &$ 0.00$& 38.5&$\hspace{-1.6mm}-18$&\hspace{1.2mm}63 & 126 & 69    & 50 & \hspace{-1.2mm}32 &0.055&$29\!-\!30$&$40.4$ & $38.5$ & $1.9$  & $0.0$ & $19\!-\!40$   &  9.3\,(0.7) & 1.06\,(11)  &  99.4 & (1) \\[1.0ex]
\multicolumn{17}{l}{\emph{(b)~~Multipole:}}\\[0.5ex]
1\,S &\hspace{-1.6mm}$-0.25$&\hspace{-1.6mm}$-0.77$& 22.8 &61 &     141 & 113 & 40 & 34 &1 &0.041&$27\!-\!28$&$17.1$ & $40.4$ &\hspace{-1.6mm}$-5.7$ &\hspace{-3.0mm}$-17.6$ & $17\!-\!35$  & 9.2\,(0.8) & 1.15\,(12) &84.6 & (2)\\
2\,N &      0.25   &\hspace{-1.6mm}$-0.77$& 22.8&\hspace{-1.6mm}$-41$&\hspace{1.2mm}39& 37 & 45 & 11 &0 &0.044&$29\!-\!30$&$28.5$ & $40.4$ &$ 5.7$ &\hspace{-3.0mm}$-17.6$ & $27\!-\!35$ & 9.2\,(0.8) & 1.14\,(12) &90.0 & (3)\\[1.5ex]
\hline\\
\end{tabular}\\[-1.0ex]
\footnotesize{(1) Unpassable access: field lines curve away from orbital plane. (2) Regular access from orbital plane. (3) Access over the rotational pole.}

\label{tab:zee}
\end{flushleft}
\end{table*}

\subsection{Stark broadening}
\label{sec:stark}

An accepted theory of Stark broadening in the presence of a magnetic
field does not exist. \citet{jordan92} opted to equate the broadening
of the individual Stark components of a Balmer line to the mean Stark
shift of all components multiplied by a factor $C\!\simeq\!0.1$ and
\citet{putneyjordan95} considered values of $C\!=\!0.1$ and 1.0 for
stars with vastly different field strengths. It is necessary,
therefore, to consider the appropriate level of the line broadening
for a given application. To this end, we adopted an approximate
post factum procedure that changes the line strengths, while
avoiding recalculation of the data base. We expressed the line
profiles in terms of an optical depth $\tau_\mathrm{\lambda}$, setting
$F_\mathrm{\lambda}\!=\!F^\mathrm{c}_\mathrm{\lambda}\,\mathrm{e}^{-\tau_\mathrm{\lambda}}$,
with $F^\mathrm{c}_\mathrm{\lambda}$ the continuum flux. We then replaced
$F_\mathrm{\lambda}$ in Eq.~\ref{eq:tomo} by
$F^\mathrm{\,new}_\mathrm{\lambda}\!=\!F^\mathrm{c}_\mathrm{\lambda}\,\mathrm{e}^{-\tau_\mathrm{\lambda}\,\eta}$
and included $\eta$ as an additional free parameter in the grid
search. For our best multipole model, we obtained $\chi^2\!=\!85.5$ at
$\eta\!=\!1$ ($C\!=\!0.1$), the best fit with
$\chi^2_\mathrm{min}\!=\!84.8$ was attained at $\eta\!=\!1.25$, and
the 90\% confidence level with $\chi^2_\mathrm{min}\!=\!87.5$ was
reached for $\eta\!=\!0.80$ and 2.50, with \chisq\ quickly rising at
still lower and higher $\eta$. Hence, our fit favors line strengths
somewhat larger than nominal ($C\!=\!0.10$, $\eta\!=\!1.0$). This
result applies to our simultaneous multi-temperature fits to spectral
flux and line strengths and may not be generally valid. We adopted
$\eta\!=\!1.25$ for the present paper and obtained the systematic
errors at the 90\% confidence level for $\eta\!=\!0.8-2.5$. The large
errors re-emphasize the need for an effort to calculate the Stark
shifts \mbox{of the individual Stark components in a magnetic field}.

\subsection{Magnetic geometry of the accreting WD}
\label{sec:magg}

The right-hand panels of Fig.~\ref{fig:cyc} show two selected magnetic
WD geometries. For simplicity, both have rotational pole, magnetic
pole, and the accretion spots on the same meridian (here the paper
plane). The secondary star is located far to the left. The footpoints
of the common magnetic axis are displaced from the respective viewing
directions by $\alpha_1$ and $\alpha_2$, with
$\alpha_1\!+\!\alpha_2\!=\!180\degr\!-\!2i$, where $i$ is the
inclination and $\alpha$ is counted positive away from and negative
toward the rotational pole.  In case~A, both spots are located between
magnetic pole and viewing direction and can accrete from the nearby
orbital plane. To reach spot\,2, the plasma must travel halfway around
the WD before it attaches to a near-polar field line. In case~B,
spot\,1 can accrete from the nearby orbital plane. Spot \,2, however,
is located between magnetic and rotational pole and the field line
leads over the rotational pole in the general direction of the
secondary star. Although energetically unfavorable, this non-standard
path may be active and it is not clear whether the trip over the pole
or the travel around the WD should be dismissed as the less likely way
to feed spot\,2.

In perusing parameter space, we found that all good fits require field
strengths larger than the spot field $B_\mathrm{sp}$ and are rather
insensitive to a lack of small field strengths.  We started from a
pure dipole model that fits the pole\,1 and pole\,2 spectra with
$B_0\!=\!37.5$ and 40.5\,MG, respectively. Adding a small quadrupole
component, leads to a common $B_0\!=\!38.5$\,MG. The parameters of
this quasi-dipole fit are listed in Table~\ref{tab:zee}.  As expected
for an inclination of 80\degr\ (Sect.~\ref{sec:system} and
Table~\ref{tab:system}), $\alpha_1\!+\!\alpha_2\!\simeq\!20$\degr,
confirming the presence of a common magnetic axis for the separately
performed fits. The colatitudes $\delta_\mathrm{f}$ of the accreting
field lines agree reasonably well with those of the 1S$-$2N geometry
in Table~\ref{tab:cyc}, considering the uncertainties of about
5\degr. Superficially, the fit seems close to perfect were it not for
the disturbing fact that the geometry probably prevents accretion in
both spots. The ribbon-like spots are offset from the respective
magnetic pole by $\vartheta_\mathrm{sp}\!\sim\!50\degr$, the field
lines in the spots reach out to only 1.7\,$R_\mathrm{wd}$, and both
field lines curve away from the orbital plane. Hence, the quasi-dipole
model provides no convincing accretion geometry.
Increasing the quadrupole component provides no remedy. For
$r_\mathrm{qua}$ up to $\pm 0.40$, none of the seemingly good fits
matches the requirements set up by the cyclotron fits. The same holds
for moderate octupole components $r_\mathrm{oct}$ up to $\pm0.40$,
some of which predict ``spots'' in the form of near-equatorial ribbons
connected by tightly closed field lines.

The situation changes fundamentally for larger octupole components
with $r_\mathrm{oct}\!\la\!-0.45$. With decreasing $r_\mathrm{oct}$,
the best-fit values of $B_0$ in the primary minima of poles 1 and 2
converge and coincide for $r_\mathrm{oct}\!=\!-0.77$ and
$r_\mathrm{qua}\!=\!\mp0.25$, respectively. Table~\ref{tab:zee} lists
the fit parameters. As required,
$\alpha_1\!+\!\alpha_2\!\simeq\!20$\degr and the spot-averaged
colatitudes $\langle \delta_\mathrm{f}\rangle$ and viewing angles
$\langle \theta_\mathrm{f}\rangle$ agree reasonably well with the
1S$-$2N cyclotron results of Table~\ref{tab:cyc}. Both spots are
located closer to the magnetic poles than in the quasi-dipole case and
the field lines are close to radial with inclinations $\langle
\beta_\mathrm{f}\rangle\!\simeq\!0$, indicating that the field lines reach
far out. The multipole model represents a convincing solution,
provided the case~B path to spot\,2 is active. In passing, we note
that enforcing case~A accretion at spot\,2 by increasing its
colatitude fails because a corresponding $\chi^2$ minimum does not
exist. Switching hemispheres, the 1N geometry is a mirror image of 1S,
but a $\chi^2$ minimum at the parameters expected for 2S does not
exist either. The general caveat holds that a multipole model with
tesseral harmonics may provide a different answer.

The two left panels of Fig.~\ref{fig:zee} show the magnetic field
distributions of the WD for the multipole model of Table~\ref{tab:zee}
at the phases of the best visibility of spots 1S and 2N.  The yellow
and green bands indicate the spot field strengths of 27 and 28 MG for
pole\,1 and 29 and 30 MG for pole\,2 and the black portions the
viewing-angle selected spots. Although both aspects belong to the same
field model, the spot geometries differ significantly. So do the full
ranges of the field strengths over the visible face of the WD (bottom
panels, see also Col.~17 of Table~\ref{tab:zee}).
The right panel of Fig.~\ref{fig:zee} shows the \mbox{pole-1} spectrum (red
curve) and the Zeeman fit (superposed black curve), which faithfully
reproduces most Zeeman lines in the spectral regions that are free of
atomic emission lines. The contributions by the spot and the
photosphere+cap are also shown individually. The Zeeman lines are
prominent in the spot component because of the small spread in field
strength, but are washed out in the photospheric component. The
fit to the pole-2 spectrum (not shown) excels at
$\lambda\!>\!4500$\,\AA, but is inferior at shorter wavelengths,
possibly because it is composed of only nine independent magnetic
spectra (Table~\ref{tab:zee}, Col.~17).

\subsection{WD parameters and their errors}
\label{sec:wdtemp}

The multipole Zeeman fit to the observed pole\,1 and pole\,2 spectra
yielded temperatures for the photosphere, cap, and spot of
$T_\mathrm{ph}\!=\!9.2\,\pm\,0.7$\,kK (Table~\ref{tab:zee}) and
$T_\mathrm{cap}\!=\!14.2\,\pm\,1.0$\,kK and
$T_\mathrm{sp}\!=\!78\,\pm\,8$\,kK (Table~10). In most fits, the flux
contributed by the cap is a minor entity. The corresponding angular
radius of the WD is
$R_\mathrm{wd}/d\!=\!\ten{(1.15\,\pm\,0.12)}{9}$\,\cmkpc. All quoted
errors refer to the 90\% confidence level and include besides the
statistical also the systematic error caused by the remaining
uncertainty in the level of Stark broadening
(Section~\ref{sec:stark}). Quadratically adding the error from an
estimated 10\% uncertainty in the flux calibration of the
spectropolarimetry gives
$R_\mathrm{wd}/d\!=\!\ten{(1.15\,\pm\,0.14)}{9}$\,\cmkpc. The error
budget of $R_\mathrm{wd}$ includes in addition the 8.1\% uncertainty
in the mean dM5--dM6 distance from Sect.~\ref{sec:vlt}, raising the
error of $R_\mathrm{wd}$ to 14.1\%. Despite its large error, the
measured radius proves helpful in determining the system parameters in
Sect.~\ref{sec:system}.

\begin{figure*}[t]

\begin{minipage}[b]{75mm}
\includegraphics[height=34.0mm,angle=0,clip]{36626f9a.ps}
\hspace{1mm}
\includegraphics[height=34.5mm,angle=0,clip]{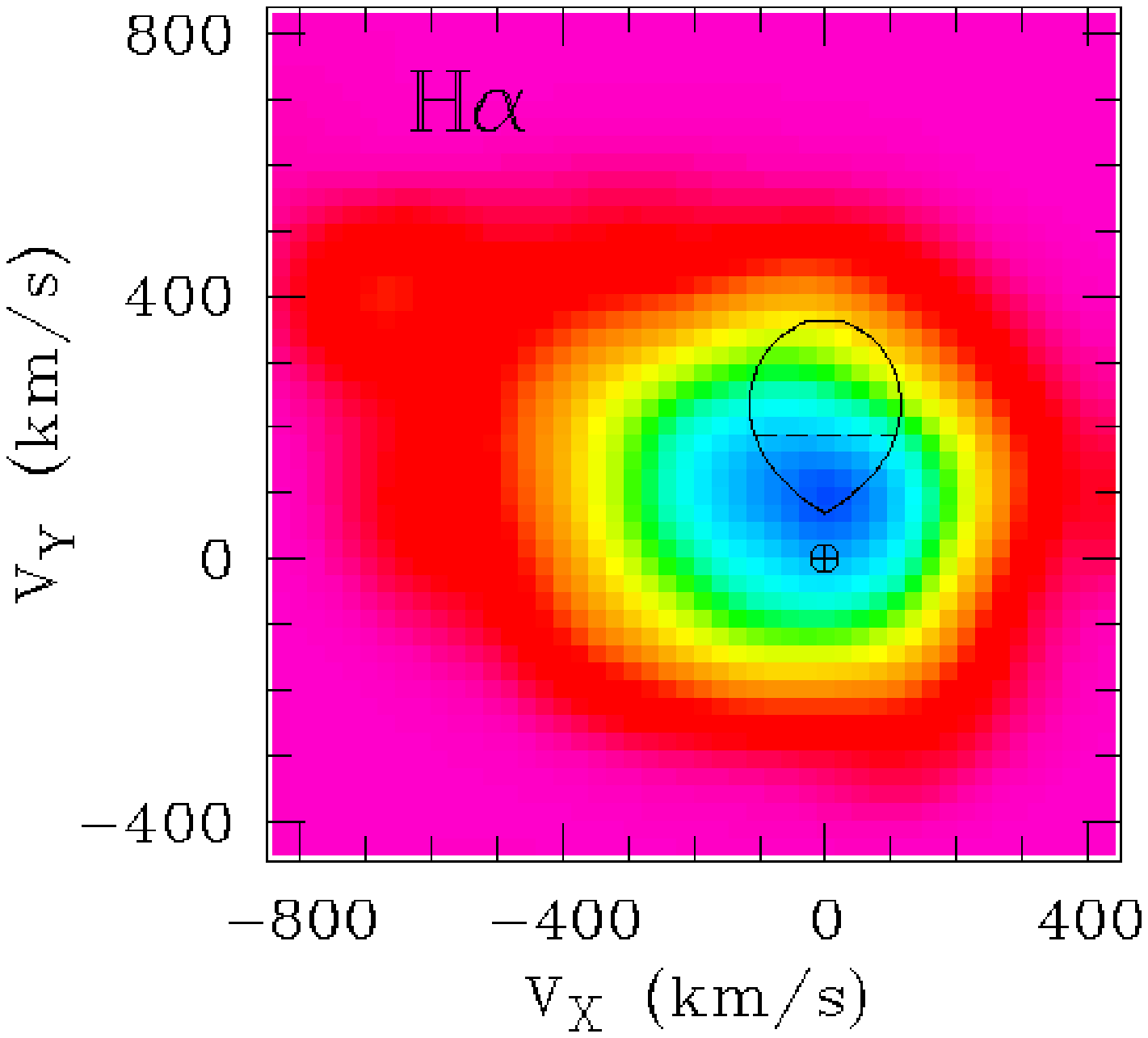}

\medskip
\includegraphics[height=34.0mm,angle=0,clip]{36626f9c.ps}
\hspace{1.6mm}
\includegraphics[height=34.5mm,angle=0,clip]{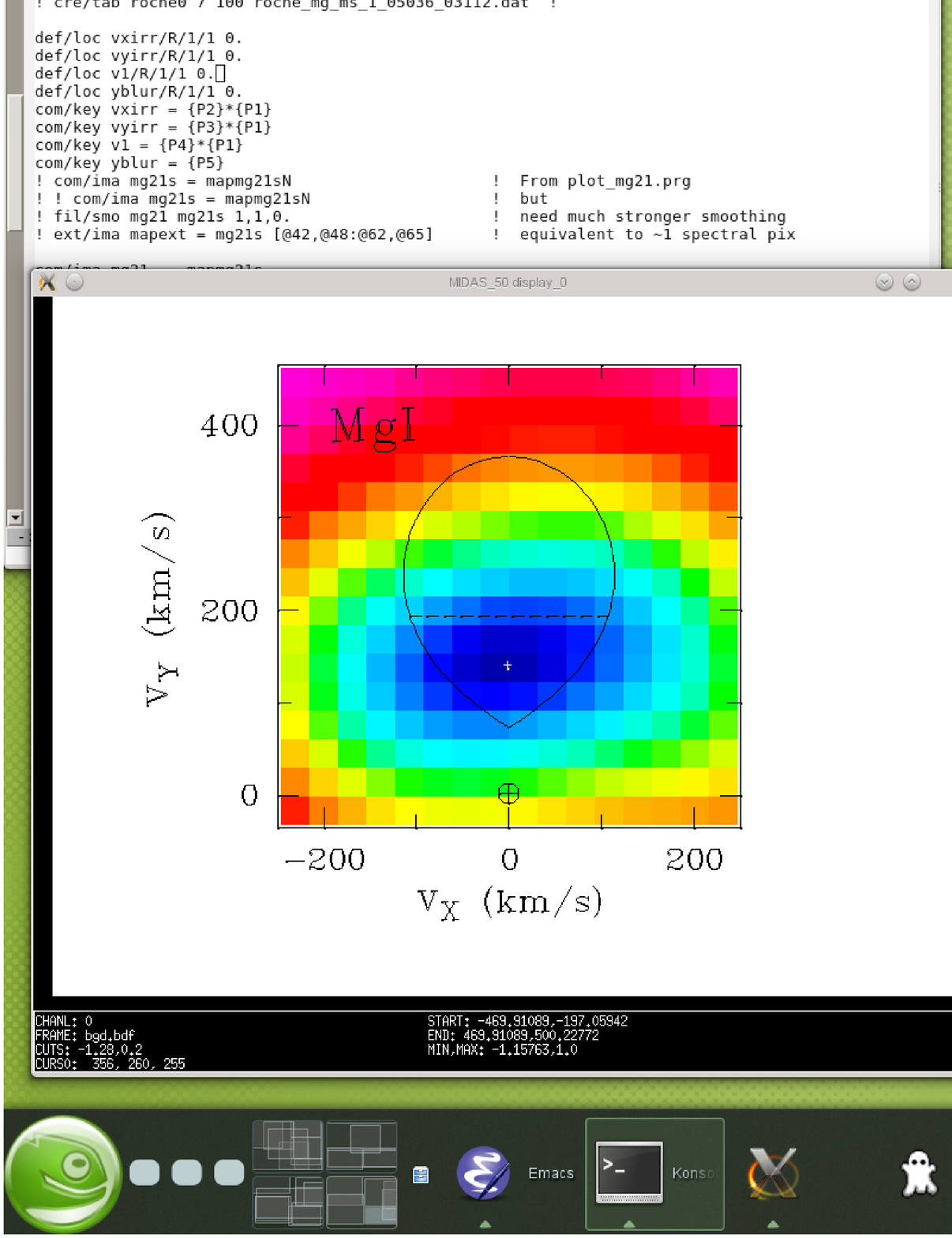}
\end{minipage}
\hspace{1mm}
\raisebox{71mm}{
\includegraphics[height=106mm,angle=270,clip]{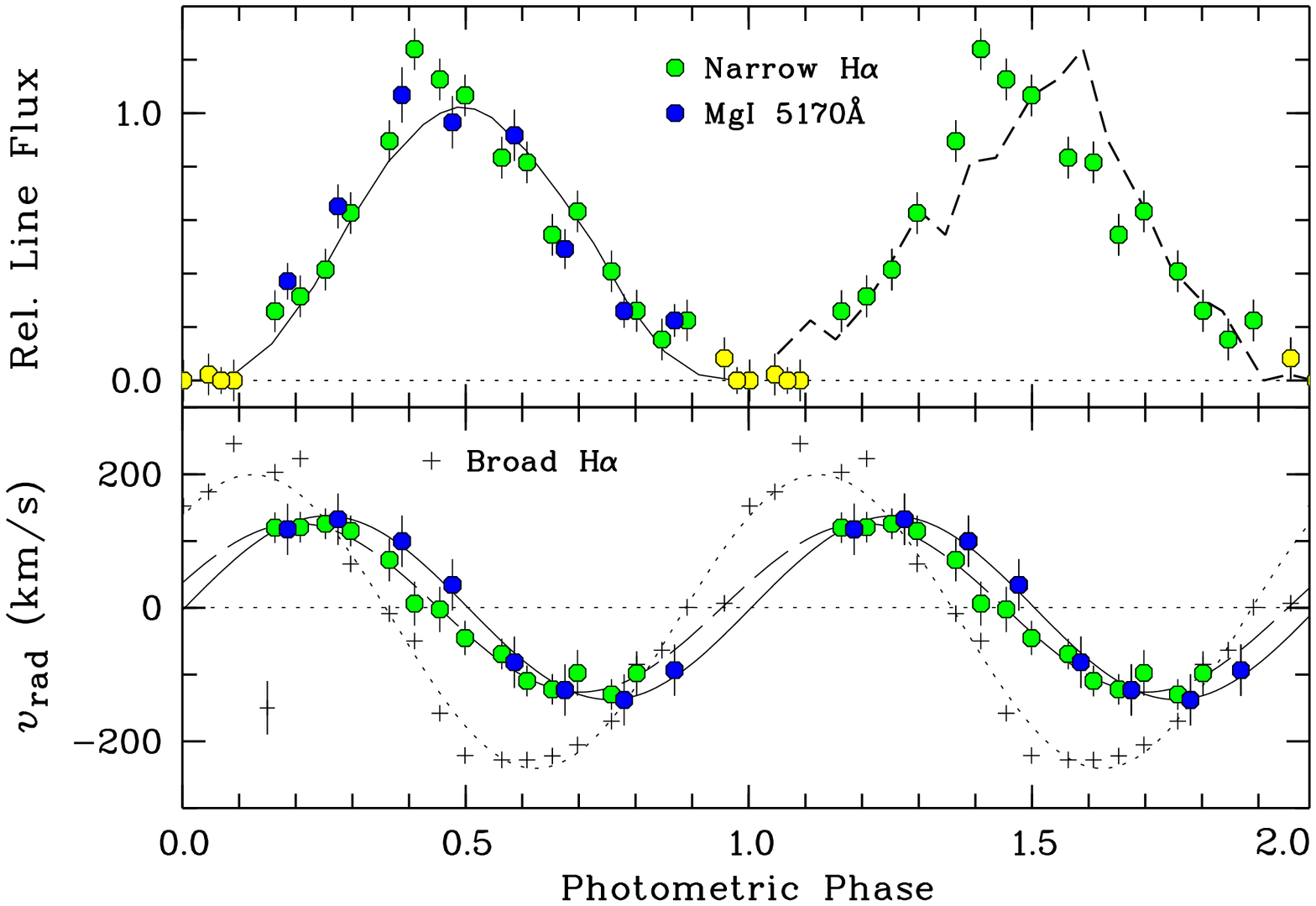}
}

\caption[chart] {\emph{Left: } \halpha\ and Mg\,I$\lambda5170$
  emission lines derived from the spectropolarimetry of 31 Dec 2008,
  with the neighboring continuum subtracted and shown twice for better
  visibility. \emph{Center: } The corresponding Doppler maps computed
  by the maximum entropy method MEM. The overlays show the Roche lobes
  of the secondary for the best bloated model of
  Sect.~\ref{sec:system}. \emph{Top right: } The first orbit shows the
  orbital flux variations of the Mg\,I line (blue) and the narrow
  component of \halpha\ (green). The second orbit shows the \halpha\
  flux and its mirror image around $\phi\!=\!0.5$ (see
  text). \emph{Bottom right: } Radial velocity curves of the MgI line (blue),
  the narrow \halpha\ component (green), and the broad \halpha\ component
  (crosses). A typical error bar for the broad line is shown in the
  lower left. }
\label{fig:vrad}
\end{figure*}

The photospheric temperature of the WD in \hyeri\ is lower than in
practically all other well-studied polars
\citep{townsleygaensicke09}. The low temperature is directly related
to the low observed equivalent width $W_\mathrm{obs}$ of the Zeeman
lines in \hyeri. The equivalent width in the theoretical Zeeman
spectra has its peak at $T_\mathrm{eff}\!=\!12$\,kK and drops rapidly
toward lower and higher temperatures. $W_\mathrm{obs}$ falls by a
factor of 3.2 below the peak value for nominal Stark
broadening. Fitting $W_\mathrm{obs}$ and the spectral slope
simultaneously, requires $T_\mathrm{ph}$ significantly below 12\,kK
and $T_\mathrm{sp}\!\gg\!12$\,kK. The fit deteriorates with rising
$T_\mathrm{ph}$ and becomes unacceptably bad at 11\,kK, even allowing
for a variation in the level of Stark broadening.

\section{System parameters}
\label{sec:sys}

\subsection{Narrow emission lines as tracers of the motion of the
secondary star}
\label{sec:nel}

Polars feature narrow emission lines of hydrogen, helium and metals
that are thought to originate on the irradiated face of secondary
star. In some polars, however, the radial velocity amplitudes of
individual lines differ. Helium lines with their high ionization
potential show lower amplitudes than hydrogen, while the
low-ionization near infrared Ca\,II lines have the greatest and rather
stable amplitudes \citep{schwopeetal00,schwopeetal11,schwopechristensen10}.
Obviously, the distribution of the emission differs between individual
lines, with the helium lines probably originating, in part, from
structures outside the chromosphere of the star, such as coronal
prominences. Modeling is straightforward as long as the emission
originates from locations geometrically close to the surface of the
star, which appears to be the case for the low-ionization metal lines
and, in some polars, for hydrogen lines.  Describing the line emission
over the secondary star requires either a dedicated theoretical model
or an empirical ansatz that derives the distribution from unfolding
highly resolved observed line profiles.

\subsection{Models of the irradiated secondary star}
\label{sec:irr} 

We calculated the radial velocity amplitude $K_2'$ of the narrow
emission line, considering a Roche-lobe filling star that is
irradiated by a source at the position of the WD. Each surface element
receives an incident flux
$f_\mathrm{in}\!\propto\!\mathrm{cos}\,\vartheta /\delta^2$, with
$\vartheta$ the angle of incidence and $\delta$ the distance from the
source. In response, it emits a line flux $f_\mathrm{line}$ that
varies with $f_\mathrm{in}$, but may depend on additional parameters.
For a given model, we calculated synthetic emission line spectra and
determined the lever arm of the line emission region
$a_\mathrm{irr}(q)\!=\!a_2\,K_2'/K_2$ as a function of the mass ratio
$q\!=\!M_2/M_1$, with $a_2\!=\!1/(1+q)$. Below, we quote polynomial
approximations for $a_\mathrm{irr}(q)$\footnote{Model BR08:
  $a_\mathrm{irr}\!=\!0.806155-0.959167\,q+0.356060\,q^2$; Model BT90:
  $a_\mathrm{irr}\!=\!0.786470-0.978076\,q+0.368182\,q^2$; both for
  $q\!=\!0.35\!-\!0.80$.}.

The irradiation model of \citet{beuermannthomas90} (henceforth BT90),
equates the line intensity emitted from a surface element to a
fraction of the total incident flux,
$f_\mathrm{line}\!\propto\!\mathrm{cos}\,\vartheta /\delta^2$. The
emitted intensity drops from a maximum at $L_1$, where
$\vartheta\!\simeq\!40$\degr, down to zero at the terminator of the
irradiated region, where $\vartheta\!=\!90$\degr. Hence, BT90 favors
emission from regions near $L_1$.  The modified version BT90m uses
$f_\mathrm{line}\!\propto\!(\mathrm{cos}\,\vartheta)^{m}/\delta^2$,
with $m$ a heuristic free parameter. This modification was motivated
by the study of irradiated WD atmospheres by \citet[][their
Sect.\,4.2]{koenigetal06}, who found that the narrow emission cores of
\lyalpha\ increased drastically when $\vartheta$ approached 90\degr\
and the incident energy was deposited increasingly higher up in the
atmosphere. There is no simple way to relate $m$ to physics,
however. An entirely different approach was taken in \citet[][their
Sects.~5.4, 6.1, and Fig.~10]{beuermannreinsch08}, where we determined
the distribution of the Ca\,II$\lambda8498$ emission as a function of
$\vartheta$ empirically by unfolding the high-resolution line profiles
of the intermediate polar EX~Hya. A parameterized form of the Ca\,II
emission model was implemented in our model BR08.  It provides an
internally consistent description of the emission of
Ca\,II$\lambda8498$ and numerous other metal lines that share the
motion of Ca\,II. The model gives larger weight to surface elements
near the terminator, equivalent to moderate limb brightening.
To put the models into perspective, we note that BR08 corresponds
approximately to BT90m with $m\!\simeq\!0.3$. BT90 and BR08 bracket
about the full range of possible irradiation scenarios of the
atmosphere of the secondary. Only models with still more pronounced
limb brightening, equivalent to BT90m with $m\,<\,0.3$, would yield a
still larger value of $a_\mathrm{irr}$ (and a still smaller primary
mass). An estimate of the remaining systematic errors is given in
Section~\ref{sec:system}.

\subsection{Narrow emission lines in \hyeri}
\label{sec:nelhy}

In its 2008 low state, \hyeri\ displayed emission lines of hydrogen,
HeI, and Mg\,I$\lambda5170$. The near infrared Ca\,II triplet was not
detectable against the cyclotron background. In \halpha, the broad
component with a FWHM of 23\,\AA\ and the unresolved narrow component
can be separated at our 10\,\AA\ spectral resolution, while this
becomes infeasible for the higher Balmer lines, which are embedded in
complex Zeeman absorption troughs. Mg\,I$\lambda5170$ is not disturbed
and is the only metal line that is sufficiently strong for a radial
velocity study. The left panels in Fig.~\ref{fig:vrad} show
pseudo-trailed spectra of \halpha\ and Mg\,I$\lambda5170$ with the
phase-dependent continuum subtracted.  The gray scale is inverted
compared with Fig.~\ref{fig:gray}. We used spectral set\,2 with 20
spectra per orbit for \halpha\ and set\,1 with 10 spectra per orbit
for the weaker Mg\,I line. The two orbits were folded and the data
shown twice for better visibility.

The upper right panel in Fig.~\ref{fig:vrad} shows the orbital flux
variations of MgI$\lambda5170$ (blue) and of the narrow \halpha\
component (green). Maximum flux occurs near $\phi\!=\!0.5$, when the
illuminated hemisphere is in view. Half an orbit later, the Mg line
disappears and the narrow \halpha\ component can no longer be
discriminated against the underlying broad component (yellow). The
upper half of the light curve is skewed, as is illustrated in the
second orbit, where the \halpha\ light curve is compared with its own
mirror image relative to $\phi\!=\!0.5$ (dashed curve). The skew is,
at least in part, due to statistical fluctuations between the two
orbits. Apart from this, the light curve is well described by the
irradiation model BR08 (solid curve). The lower right panel of
Fig.~\ref{fig:vrad} shows the radial velocity curves of
MgI$\lambda5170$ (blue), of the narrow component of \halpha\ (green),
and of the broad \halpha\ emission (crosses). Although
MgI$\lambda5170$ is weaker than \halpha, the radial velocities have
similar errors because the former were derived from single-Gaussian
and the latter from the more uncertain double-Gaussian
fits. MgI$\lambda5170$ has a radial velocity amplitude of
$K_2'\!=\!139\pm10$\,\kms\ with a blue-to-red zero-crossing phase of
$\phi_0\!=\!0.02\pm0.02$, the narrow \halpha\ line has
$K_2'\!=\!125\pm9$\,\kms\ with $\phi_0\!=\!-0.06\pm0.02$. The broad
component has a velocity amplitude $K_\mathrm{broad}\!=\!220
\pm14$\,\kms with $\phi_0\!=\!-0.13\pm0.02$ and
$\gamma\!=\!-21\pm9$\,\kms\ relative to the narrow component. All
errors refer to the 90\% confidence level.

We investigated the origin of the lines by calculating Doppler
tomograms, using the maximum entropy method MEM
\citep{spruit98,marshschwope16}. Given the small number of phase
intervals, the tomograms are sensitive to noise and have been slightly
smoothed with a velocity filter corresponding to 0.3 spectral
resolution elements. The resulting tomograms are shown in the
center panels of Fig.~\ref{fig:vrad}, with the outline of the Roche
lobe of the secondary for our best-fitting dynamical model overplotted
(Table~\ref{tab:system}, line~4). The bulk of the emission can be
uniquely allocated to the illuminated face of the secondary star and
the vicinity of the inner Lagrangian point $L_1$. The rainbow color
scale ranges from blue for the highest intensity down to red.
Overall, the \halpha\ tomogram is tilted toward the leading hemisphere
of the secondary, with the asymmetry related to the finite negative
$\phi_0$ and the existence of the underlying broad component. The
latter with its best visibility at $\phi\!=\!0.60$ is represented by
the tail that extends to $V_\mathrm{X}\!=\!-700$\,\kms\ and
$V_\mathrm{Y}\,=\!+450$\,\kms. This direction differs significantly
from that of the standard ballistic stream seen in many polars in
their high states, which moves in velocity space from $L_1$ to large
negative $V_\mathrm{X}$ at nearly constant $V_\mathrm{Y}$.  A tail at
similarly odd velocities was seen in the He\,II$\lambda4686$ tomogram
of AM~Her \citep{staudeetal04} and tentatively interpreted in terms of
a non-standard accretion stream that couples from the secondary
immediately to a polar field line of the WD. This is an intriguing
proposition in view of our suggestion in Sect.~\ref{sec:magg} that
pole~2 of \hyeri\ is fed by such a scenario. In view of these
idiosyncrasies, we should be wary of interpreting the narrow \halpha\
line in \hyeri\ as of purely chromospheric origin.

The Mg\,I$\lambda5170$ line is free of the complications by a broad
component, the tomogram looks more regular, and the phase of zero
radial velocity is consistent with inferior or superior conjunction of
the secondary star.  The bottom center panel of Fig.~\ref{fig:vrad}
shows the enlarged central portion of the tomogram. The emission is
centered on the illuminated part of the star, with the peak intensity
occurring at a $Y$-velocity that agrees with
$K_2'\!=\!139\!\pm\!10$\,\kms\ obtained from the radial-velocity
analysis (small white cross).

\begin{table*}[t]
\begin{flushleft}
  \caption{System parameters of \hyeri\ for models with main sequence
    and bloated secondary stars, derived with the radial velocity
    amplitude of the Mg\,I line and the irradiation model BT08. The
    columns contain $K_2'$ (2), the bloating factor $f_3$ (3),the
    derived model parameters (3--11), the model radius of the He-core
    of CO-core WD of the mass of Col.~(11) cooled to 10\,kK (12,13),
    the distance (14), the spectroscopically derived WD radius (15),
    the ratio of the radii (16), the angular radius of the WD (17),
    and its implied photospheric temperature (18).}

\medskip
\begin{tabular}{@{\hspace{1.0mm}}c@{\hspace{-1.0mm}}c@{\hspace{0.0mm}}c@{\hspace{3.0mm}}c@{\hspace{3.0mm}}c@{\hspace{3.0mm}}c@{\hspace{3.0mm}}c@{\hspace{1.0mm}}c@{\hspace{2.0mm}}c@{\hspace{2.0mm}}c@{\hspace{2.0mm}}c@{\hspace{2.0mm}}c@{\hspace{1.0mm}}c@{\hspace{2.0mm}}c@{\hspace{3.0mm}}c@{\hspace{1.0mm}}c@{\hspace{0.0mm}}c@{\hspace{1.0mm}}c}\\[-5ex]
\hline\hline \\[-1.5ex]
(1) & (2)    & (3)   & (4) & (5) &(6) & (7) & (8)       & (9)   &  (10) & (11) &   (12)      & (13)            & (14) & (15) & (16)  & (17) & (18)  \\
Row&$K_2^{'}$& $f_3$ & $i$ & $q$ &$M_2$&$R_2$&$K_2/K_2^{'}$ & $K_\mathrm{L1}$ & $K_\mathrm{term}$ & $M_1$ & $R_\mathrm{1,He}$ & $R_\mathrm{1,CO}$ & $d$ & $R_\mathrm{1,sp}$ & $R_\mathrm{1}/R_\mathrm{1,sp}$ & $R_1/d$ & $T_\mathrm{ph}$ \\
  &(\kms)  &   & (\degr) &   &(\msun)&(\rsun)&         & \multicolumn{2}{c}{(\kms)}   &(\msun)& \multicolumn{2}{c}{($10^9$\,cm)}  &(pc) & ($10^9$\,cm) & & ($10^9\!$cm/kpc) & (kK)\\[0.5ex]  
\hline\\[-1ex]                                                                                                                                                
\multicolumn{18}{l}{\emph{ (a) main sequence models:} }\\ [0.5ex]                                                                                                                
 1 & 129 & 1.000 & 79.1 & 0.656 & 0.308 & 0.320 & 1.83 & 57 & 186 & 0.469 & 1.116 &       & 1167$\,\pm\,$95 & 1.343$\,\pm\,$0.189 & 0.83 & 0.956 & 10.19  \\                      
 2 & 139 & 1.000 & 79.8 & 0.615 & 0.305 & 0.318 & 1.76 & 67 & 196 & 0.497 & 1.065 &       & 1160$\,\pm\,$94 & 1.334$\,\pm\,$0.188 & 0.80 & 0.918 & 10.42  \\                      
 3 & 149 & 1.000 & 80.4 & 0.577 & 0.304 & 0.317 & 1.71 & 76 & 207 & 0.526 &       & 0.969 & 1154$\,\pm\,$93 & 1.327$\,\pm\,$0.187 & 0.73 & 0.840 & 10.92  \\[1.0ex]               
\multicolumn{18}{l}{\emph{ (b) Best-fit model with bloated secondary star}}          \\   [0.5ex]                                                                                 
 4 & 139 & 1.120 & 80.5 & 0.568 & 0.235 & 0.290 & 1.70 & 72 & 192 & 0.413 & 1.217 &       & 1054$\,\pm\,$85 & 1.217$\,\pm\,$0.172 & 1.00 & 1.150 &\hspace{1.4mm}9.20\\[1.0ex]     
\multicolumn{18}{l}{\emph{ (c) Models with bloated secondaries, delineating the 90\% confidence limits} }\\    [0.5ex]                                                           
 5 & 139 & 1.035 & 80.0 & 0.600 & 0.281 & 0.309 & 1.74 & 68 & 195 & 0.469 & 1.117 &       & 1127$\,\pm\,$91 & 1.296$\,\pm\,$0.183 & 0.86 & 0.991 & 10.00  \\                     
 6 & 149 & 1.135 & 81.4 & 0.524 & 0.224 & 0.285 & 1.64 & 83 & 202 & 0.427 & 1.190 &       & 1037$\,\pm\,$84 & 1.192$\,\pm\,$0.168 & 1.00 & 1.148 &\hspace{1.4mm}9.20 \\                     
 7 & 129 & 1.095 & 79.7 & 0.617 & 0.247 & 0.297 & 1.77 & 62 & 183 & 0.401 & 1.244 &       & 1081$\,\pm\,$88 & 1.244$\,\pm\,$0.175 & 1.00 & 1.150 &\hspace{1.4mm}9.20\\                     
 8 & 139 & 1.190 & 81.1 & 0.541 & 0.201 & 0.275 & 1.66 & 75 & 189 & 0.371 & 1.313 &       & 1002$\,\pm\,$81 & 1.152$\,\pm\,$0.162 & 1.14 & 1.310 &\hspace{1.4mm}8.60  \\[1.0ex]              
 \hline\\                                                                                                                                                                                              
\end{tabular}\\                                                                                                        
\label{tab:system}                                                                                                                                                                                   
\end{flushleft}
                                              
\vspace{-7mm}
\end{table*}

\subsection{Mass-radius relation of the secondary star}
\label{sec:massrad}

Deriving stellar masses requires that we adopt a mass-radius relation
$R_2(M_2)$ for the Roche-lobe filling secondary star. We used
theoretical models by \citet[][henceforth BCAH and
BHAC]{baraffeetal98,baraffeetal15} for main sequence stars of
solar composition evolved to 1\,Gyr. This is the approximate cooling
age of the WD in \hyeri, discounting compressional heating by
accretion, and the minimal age of the secondary. For ease of
application, we represented the radii by power laws
$R_2(M_2)$\,\footnote{Baraffe et al. (2015) main sequence mass-radius
  relations, $M_2/M_\odot\!= \!0.15\!-\!0.40$\,\msun, $MH\!=\!0$,
  1\,Gyr: $R_2/R_\odot\!=\!0.736(M_2/M_\odot)^{0.761}$; 5\,Gyr:
  $R_2/R_\odot\!=\!0.766(M_2/M_\odot)^{0.780}$; 10\,Gyr:
  $R_2/R_\odot\!=\!0.787(M_2/M_\odot)^{0.792}$; $MH\!=\!-1$, 1\,Gyr:
  $R_2/R_\odot\!=\!0.735(M_2/M_\odot)^{0.793}$.}  Since secondary
stars in CVs are known to be more or less bloated compared with field
stars, we considered radii expanded over those of the BHAC models by
the following processes: (i) magnetic activity and spot coverage
\citep{chabrieretal07,moralesetal10,kniggeetal11,parsonsetal18}; (ii)
tidal and rotational deformation of Roche-lobe filling stars
\citep{renvoizeetal02}; and (iii) inflation by magnetic braking that
drives the star out of thermal equilibrium \citep{kniggeetal11}.
Effect (i) describes the radius excess that compensates for the
reduced radiative efficiency caused by starspots. \citet{kniggeetal11}
and \citet{parsonsetal18} found a mean excess of 5\% for stars with
mass below $0.35$\,\msun. \citet{moralesetal10} and
\citet{kniggeetal11} argued that high-latitude spots may mimic a
larger radius in certain eclipsing binaries, accounting for 3\% of the
excess. Proceeding conservatively, we accept this argument and adopted
a bloating factor $f_1\!=\!1.020$. A Roche-lobe filling star in a
short period binary can not escape effect (ii), which increases the
radius by a factor $f_2\!=\!1.045$ independent of $q$
\citep{renvoizeetal02,kniggeetal11}.  Effect (iii) is described by a
free factor $f_\mathrm{3}$ that may range from unity up to about
1.30. The adopted stellar radii $R_2\!=\!f_1f_2f_3\,R_\mathrm{BHAC}$
are fully consistent with those employed by \citet{kniggeetal11} in
their evolutionary sequences. Using the models of stars with solar
composition evolved to ages of 5 or 10~Gyr instead of 1~Gyr, the
dynamical solution yields WD masses lower by 3\% or 5\%,
respectively. For a metal-poor secondary with $[M/H]\!=\!-1$, the
masses would be higher by 8\% at 1 Gyr, but correspondingly lower
again for larger ages.

\subsection{Component masses and distance}
\label{sec:system}

\begin{figure}[t]

\medskip
\includegraphics[height=89.0mm,angle=270,clip]{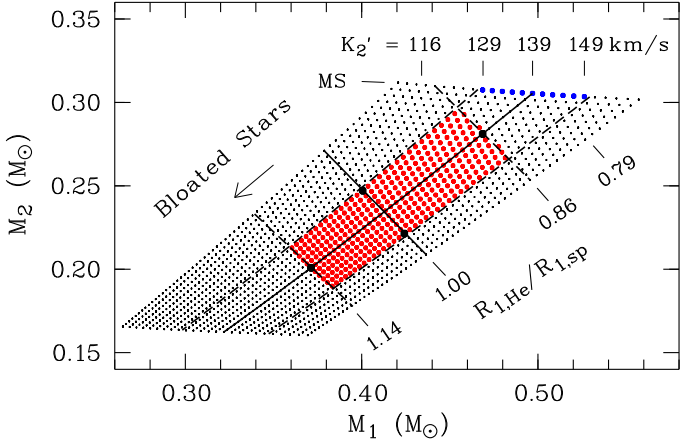}
\caption[chart] {Dynamic models of \hyeri\ in the $M_1\!-\!M_2$
  plane. The irradiation model is BR08, $K_2'$ ranges from 111 to 159
  \kms\ in steps of 2\,\kms, and the bloating factor $f_3$ ranges from
  1.000 to 1.300 in steps of 0.005. The models with main sequence (MS)
  secondaries are marked by blue dots. Red dots denote the models that
  match both, the measured radial-velocity amplitude of
  Mg\,I$\lambda5170$ and the spectroscopically measured WD radius
  within their 90\% confidence errors.}
\label{fig:m1m2} 

%\vspace{-3mm}
\end{figure}

For a given irradiation model and a mass-radius relation of the
secondary star, we obtained the system parameters that match the
radial-velocity amplitude $K_2'$ and the eclipse duration
$\Delta\,t_\mathrm{ecl}$. We adopted $K_2'\!=\!139\!\pm\!10$\,\kms\ of
the MgI$\lambda5170$ line and BR08 as the standard. Results are
presented in Table~\ref{tab:system} and Fig.~\ref{fig:m1m2}. The
derived parameters include the masses and radii of the components, the
distance $d$ obtained from the angular radius of the secondary star
(Sect.~\ref{sec:vlt}) and the spectroscopic radius $R_\mathrm{1,sp}$
of the WD obtained from $d$ and the angular radius of the WD
(Sect.~\ref{sec:wdtemp}). The listed model radii in Cols. (12) and
(13) of Table~\ref{tab:system} refer to He-core and CO-core WDs with
thick hydrogen envelopes\,\footnote{The envelope mass varies between
  $M_\mathrm{H}\!=\!\ten{8}{-5}$ at $M_1\!=\!0.52$\,\msun\ and
  $\ten{3}{-4}$\,\msun\ at $0.3$\,\msun, staying below the respective
  ignition masses.} evolved to $T_\mathrm{eff}\!=\!10$\,kK
\citep{paneietal07,renedoetal10,althausetal13}\footnote{http://evolgroup.fcaglp.unlp.ar/TRACKS/DA.html}.
The radii for an effective temperature of 9\,kK are only minimally
smaller.

To start with, we take the secondary to be an unbloated main sequence
star with $f_3\!=\!1.0$, an assumption that yields the maximum primary
mass, but disregards the spectroscopic information from
Section~\ref{sec:wdtemp}.  For the Mg\, line with
$K_2'\!=\!139\!\pm\!10$\,\kms\ and BR08, the component masses are
$M_1\!=\!0.497\!\pm\!0.029$\,\msun\ and
$M_2\!=\!0.305\!\pm\!0.002$\,\msun\ (Table~\ref{tab:system}, lines 1
to 3, 90\% confidence errors). Interestingly, the \halpha\ value
$K_2'\!=\!125\!\pm\!9$\,\kms\ and BT90, yield practically the same
primary mass, $M_1\!=\!0.487\!\pm\!0.026$\,\msun, but this combination
lacks the internal consistency that exists between the Mg\,I line and
the metal-line calibrated model BR08. Cross-combining velocity
amplitude and irradiation model gives an indication of the remaining
systematic error: \halpha\ and BR08 give $M_1\!=\!0.458$\,\msun, Mg\,I
and BT90 give $M_1\!=\!0.528$\,\msun, or combined
$M_1\!=\!0.493\!\pm\!0.035$\,\msun. Hence the so-defined systematic
error is of the same size as the statistical error of 0.029\,\msun. In
summary, the assumption of a main sequence secondary identifies the
primary either as a He-core WD or a CO-core WD very close to its
minimum mass of $0.53\pm0.02$\,\msun\
\citep{moehleretal04,kaliraietal08}.
The fault with the main sequence assumption is the neglect of the spectroscopic
evidence of Section~\ref{sec:wdtemp} on the angular radius and the
effective temperature of the WD. The implied model radius of the
primary in either Col.~(12) or Col.~(13) of Table~\ref{tab:system},
lines $1\!-\!3$, falls far short of the spectroscopically determined
radius $R_\mathrm{1,sp}$ in Col.~(15), calculated from
$R_\mathrm{wd}/d\!=\!\ten{(1.15\,\pm\,0.14)}{9}$\,\cmkpc\
(Sect.~\ref{sec:wdtemp}) and the $R_2$-dependent distance $d$ in
Col.~(14), where we have added the errors quadratically. The
employed angular radius of the WD belongs to the best-fit photospheric
temperature $T_\mathrm{ph}\!=\!9.2\,\pm\,0.7$\,kK. Had the WD the
radius of Col.~(12) or (13), the observed spectral flux would demand
that its temperature would be that in Col.~(18). As noted in
Section~\ref{sec:wdtemp}, a decent Zeeman spectral fit can not be
achieved for $T_\mathrm{ph}\!>\!10$\,kK, further reducing the
probability that the primary in \hyeri\ is a low-mass CO WD.

In a second step, we considered models with bloated secondary
stars. We calculated a grid of models with radial velocity amplitudes
$K_2'\!=\!110$ to 160\,\kms\ and expansion factors $f_3\!=\!1.0$ to
1.3 in steps of 2\,\kms\ and 0.005, respectively. We identified $K_2'$
with the radial velocity amplitude of the Mg\,I$\lambda5170$ line and
converted it to $K_2$, using the BR08 model. The resulting component
masses are depicted in Fig.~\ref{fig:m1m2}, where each model is
represented by a dot. The main sequence models considered above are
marked in blue. Models that comply with the Mg\,I amplitude
$K_2'\!=\!139\!\pm\!10$\,\kms, are located between the two dashed
lines, extending from the upper right to the lower left. Along this
path, the bloating factor $f_3$ of the secondary star increases, the
mass of the WD decreases, its radius increases, and its temperature
decreases. Models with He-core WDs, whose radii agree within the
uncertainties with the spectroscopically determined WD radius,
$R_\mathrm{1,He}/R_\mathrm{1,sp}\!=\!1.00\!\pm\!0.14$ (90\% confidence
error, Section~\ref{sec:wdtemp}) are located between the two dashed
lines that run from the upper left to the lower right, and models that
match both conditions are marked by red dots. The optimal dynamical
model in line 4 of Table~\ref{tab:system} corresponds to the
intersection of the two solid lines and the WD parameters at this
point correspond to those of the optimal multipole Zeeman fit in
Table~\ref{tab:zee}.
Lines $5\!-\!8$ of Table~\ref{tab:system} contain the model parameters
for the four cardinal points of the red-dotted region, marked by the
four black dots. Col.~(3) of the table lists the bloating factor
$f_3$. Bloating ranges from a minimal 3.5\% to 19\%, indicating that
the secondary is only moderately expanded as may be expected for a
polar that experiences reduced magnetic braking
\citep{WickramasingheWu94,webbinkwickramasinghe02}. As discussed in
Section~\ref{sec:disc}, minimal bloating that goes along with a low
accretion rate is also required to explain the low photospheric
temperature of the WD in terms of compressional heating
\citep{townsleygaensicke09}. At the optimal position and at two of
the cardinal points, the model and observed WD radii in Cols.~(12) and
(15) agree. At the two other cardinal points, they disagree by the
permitted $\pm\!14$\% (column 16). The 90\% confidence region for the
combined dynamical and spectroscopic fit is defined by a quasi-ellipse
that is inscribed to the red-dotted quadrilateral and passes through
the cardinal points (not shown). It limits the component masses to
$M_1\!=\!0.413\,+\!0.058,-0.044$\,\msun\ and
$M_2\!=\!0.235\!+\!0.048,-0.038$\,\msun\ (90\% confidence errors), or
$M_1\!=\!0.42\!\pm\!0.05$\,\msun\ and
$M_2\!=\!0.24\!\pm\!0.04$\,\msun, with $q$ between 0.51 and 0.63. The
combined dynamical and spectroscopic fit identifies the primary in
\hyeri\ as a low-mass WD, consistent with having a helium core. The
mass of the secondary is normal for a CV with an orbital period of
2.855\,h. It may be fully convective or retain a radiative core,
depending on its prehistory \citep{kniggeetal11}.
  
\begin{table}[t]
\begin{flushleft}
\caption{System parameters of \hyeri\ with 90\% confidence errors.}
\begin{tabular}{@{\hspace{2.0mm}}l@{\hspace{6mm}}rl@{\hspace{3mm}}r}
\hline\hline \\[-1.0ex]
Parameter                              & Value           && Error            \\ [0.5ex]
\hline\\[-0.5ex]
Orbital period \porb\ (s)              & 10278.9            &&                  \\
Eclipse duration FWHM $\Delta\,t_\mathrm{ecl}$ (s)       &  910.6             &&    1.5              \\[0.5ex]
Primary mass $M_1$ (\msun)             & 0.42           && $0.05$             \\
Primary radius $R_1$ at $T_\mathrm{ph}$ ($10^{9}$\,cm)& 1.22  && $0.10$       \\
Secondary mass $M_2$ (\msun)           & 0.24            && $0.04$             \\
Secondary radius $R_2$ (\rsun)         & 0.29            && $0.02$             \\
Mass ratio $q$                         & 0.57            && $0.06$             \\
Binary separation $a$ ($10^{10}$\,cm)   & 6.16             && $0.28$              \\
Inclination $i$ (\degr)                & 80.6             &&  $0.9$              \\[0.5ex]
Distance $d$ (pc)                      & 1050             && $110$               \\[0.5ex]
$T_\mathrm{eff}$ of secondary (K) &\hspace{-1.0mm}3000      && $100$         \\
Luminosity of secondary  $L_\mathrm{bol}$ ($10^{31}$\,erg/s) &\hspace{-1.0mm}2.3      && \hspace{-0.3mm}0.6         \\[0.5ex]
$T_\mathrm{eff}$ of photosphere of WD $T_\mathrm{ph}$ (kK) &9.2 && \hspace{-0.3mm}0.7 \\
$T_\mathrm{eff}$ of pole cap of WD $T_\mathrm{cap}$ (kK)  &14.2 && \hspace{-0.3mm}1.0 \\
Weighted mean of $T_\mathrm{ph}$ and $T_\mathrm{cap}$ (kK)  &10.1 && \hspace{-0.3mm}0.9 \\
$T_\mathrm{eff}$ of spot\,1 of WD $T_\mathrm{sp}$ (kK)     & 78.0& \hspace{-3.5mm}$^{(1)}$ & \hspace{-0.0mm}8.0 \\[0.5ex]
Magnetic field strength in spot1 $B_\mathrm{sp,1}$~~(MG) & 27.4           && 0.2 \\
Magnetic field strength in spot2 $B_\mathrm{sp,2}$~~(MG) & 28.7           && 0.3 \\[1.0ex]
\hline \\

\end{tabular}\\[-1.0ex]
\footnotesize{(1) Subject to the spot size.}
\label{tab:sum}
\end{flushleft}

\vspace{-6mm}
\end{table}

The observed parameters of \hyeri\ are summarized in Tab.~10.
At a distance $d\!=\!1050\,\pm\,110$\,pc and a galactic latitude
$b\!=\!-26.1$\degr, it is located close to 500\,pc below the galactic
plane. The second Gaia data release \citep{gaia18a} yielded a parallax
$\pi\!=\!-0.1\!\pm\!0.8$\,mas at a mean $g\,=\,20.3$, or a distance of
$d\!>\!830$\,pc \citep{bailerjonesetal18}, consistent with all entries
in Table~\ref{tab:system}.

\subsection{Luminosity and accretion rate}
\label{sec:sed}
 
\begin{figure}[t]
\includegraphics[height=89mm,angle=270,clip]{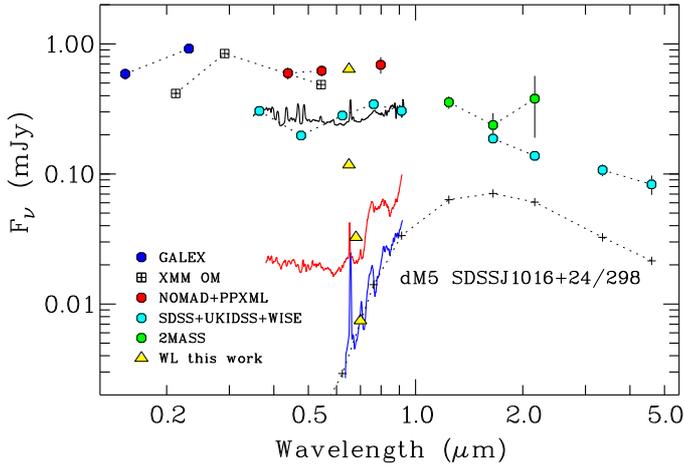}
\caption[chart]{Spectral energy distribution of \hyeri\ from the far 
 ultraviolet to the near infrared based in part on publicly available 
 photometric data accessed via the Vizier SED tool available at the 
 CDS in Strasbourg.}
\label{fig:sed}

\end{figure}

The spectral energy distribution (SED) in Fig.~\ref{fig:sed} provides
an overview of the long-term variability of \hyeri. It shows the
spectrum of 1993 from Fig.~\ref{fig:spec} (black curve), that of 2008
from Fig.~\ref{fig:cyc} (red), and the eclipse spectrum from the
center panel of Fig.~\ref{fig:ecl} (blue). The photometric data
were accessed via the Vizier SED tool provided by the
CDS\footnote{Centre de Donn\'ees astronomiques de Strasbourg,
  http://vizier.u-strasbg.fr/vizier/sed/}. They include the
SDSS\footnote{Data Release 15, http://www.sdss.org/dr15}, the UKIDSS,
and the Wide-field Infrared Survey \citep[WISE;][]{cutrietal13} as
cyan blue dots, the Two Micron All Sky Survey
\citep[2MASS;][]{skrutskietal06} as green dots, the Galaxy Evolution
Explorer \citep[GALEX;][]{martinetal05,bianchietal11} as blue dots,
and the catalogs PPMXL \citep{roeseretal10} and NOMAD
\citep{zachariasetal05} as red dots. The range of the MONET WL
measurements of Fig.~\ref{fig:lc} is indicated by the four open
triangles that refer to the observations of January 2010, November
2010, 2017/18, and the remnant flux inside the eclipse. They delineate
the same range of flux levels as the independent photometric and
spectrophotometric observations. The secondary star (+) represents a
lower flux limit in the red part of the SED.

To a first approximation, the high state is characterized by a roughly
flat SED from the infrared to the FUV, with
$F_\mathrm{\nu}\!\simeq\!0.5$\,mJy and an integrated energy flux of
$F_\mathrm{UV,opt}\!\simeq\!\ten{1.5}{-11}$\,\ergs. At a distance of
1\,kpc, the high-state UV--optical luminosity amounts to
$L_\mathrm{UV,opt}\!\simeq\!\ten{2.0}{33}$\,\erg\ for emission into
4$\pi$. The 2002 UV and visual photometry with the XMM-Newton optical
monitor demonstrates that the simultaneous X-ray observation was also
taken in a near-high state.
The bolometric X-ray flux measured in the ROSAT and XMM-Newton
observations is not well established. In Sect.~\ref{sec:xray} we
quoted a conservative high-state level of the bolometric soft X-ray
flux of $F_\mathrm{X}\!\simeq\!\ten{1.5}{-11}$\,\ergs, although the
X-ray flux may have been significantly higher at times.  Hence, as a
conservative estimate the total high-state luminosity, not accounting
for the hard X-ray region, was
$L_\mathrm{X,UV,opt}\!\simeq\!\ten{4}{33}$\,\erg. For the best-fit
model in Table~\ref{tab:system}, line 4 and Table~10 with a WD of
0.42\,\msun, this luminosity requires an accretion rate of
$\dot{M}\!\simeq\!\ten{9}{16}$\,g\,s$^{-1}\!=\!\ten{1.4}{-9}\,$\msun\,yr$^{-1}$,
On the other hand, the integrated orbital mean cyclotron flux in the
2008 low state, including the extrapolation into the unobserved
infrared regime, amounted to
$F_\mathrm{opt,IR}\!\simeq\!\ten{3}{-14}$\,\ergs. The implied remnant
accretion rate was a mere
$\dot{M}\!\simeq\!\ten{1.4}{-12}\,$\msun\,yr$^{-1}$, three orders of
magnitude lower than in the high state.

\section{Discussion}
\label{sec:disc}

\hyeri\ belongs to the rather small group of about a dozen eclipsing
polars that may be considered well studied \citep[][final edition
7.24, 2016]{ritterkolb03}.  When looked at superficially, it is a rather
uninspiring member of the class, being distant and faint, and lapsing
into prolonged states of low accretion\footnote{A long-term light
  curve is available from the Catalina Sky Survey at
  http://nesssi.cacr.caltech.edu/catalina/CVservice/CVtable.html}.
The present study, however, reveals \hyeri\ as a system that is
peculiar in several respects and in part unique among polars. 

We identified it as a permanent two-pole accretor with both poles
being active for accretion rates differing by three orders of
magnitude.  Two nearly opposite accretion spots with similar field
strengths seem to suggest a dipolar field, but that is probably an
illusion. Our combined quasi-dipole fit to the spectra of both poles
is formally good and yields field directions in both accretion spots
that are consistent with the results of our cyclotron line fits, but
nevertheless the magnetic geometry is characterized by tightly closed
field lines in both spots that seem to preclude accretion. In an
extensive grid search, we discovered an alternative close-to-perfect
fit for a field structure with a polar field strength of the dipole
$B_\mathrm{dip}\!=\!40.4$\,MG and relative field strengths of
quadrupole and octupole $B_\mathrm{qua}/B_\mathrm{dip}\!=\!\mp\,0.14$
and $B_\mathrm{oct}/B_\mathrm{dip}\!=\!-\,0.44$. It possesses open
field lines in both spots, allowing the southern spot\,1 to accrete
from the nearby orbital plane, while the field in the northern spot\,2
leads over the rotational pole to a place somewhere near the secondary
star. If that path is active, it obviates the need for the plasma
accreted at spot\,2 to travel halfway around the WD until it attaches to
a near-polar field line.  We caution, however, that models that
include the full zoo of tesseral harmonics may favor still different
magnetic geometries. Furthermore, Zeeman fits based on intensity
spectra alone may not yield a unique result. The simultaneous fit to
intensity and circular polarization spectra offers better perspectives
as demonstrated by \citet{euchneretal02,euchneretal05,euchneretal06}
and \citet{beuermannetal07}.

The Zeeman spectra of 2008 were obtained in a state of low accretion and
yielded a mean mass-flow rate $\dot m$ close to that of the
bombardment solution of \citet{woelkbeuermann92}. The spot fields
obtained from the Zeeman and cyclotron fits differ by about 3\%,
suggesting that the cyclotron emission originates at a height above
the photosphere of roughly 0.01\,\rwd, provided both originate at the
same position on the WD. This altitude is in the same ballpark as the
shock height of $\sim\!0.006$\,\rwd\ predicted by two fluid cooling
theory \citep{fischerbeuermann01} for the mass flow rate listed in
Table~\ref{tab:cyc}.

The property by which \hyeri\ deviates most drastically from other
polars, is the low primary mass of
$M_\mathrm{wd}\!=\!0.42\!\pm\!0.05$\,\msun, based on the simultaneous
dynamical and Zeeman spectral analysis. The dynamical analysis alone
limits the primary mass to $M_1\!<\!0.53$\,\msun. The secondary mass
of $M_2\!=\!0.24\!\pm\!0.04$\,\msun\ appears normal for a CV of
2.855\,h orbital period. Only two other polars in the Ritter\,\&\,Kolb
catalog \citep[][final edition 7.24, 2016]{ritterkolb03} have reported
WD masses below 0.50\,\msun. \citet{gaensickeetal00} derived the
angular radius of the WD in V1043 Cen from far-ultraviolet
spectroscopy and the radius and mass from a distance estimate of
200\,pc. At the Gaia distance of 172\,pc \citep{gaia18a}, however, the
radius of the WD is $\ten{9.5}{8}$\,cm and, at a temperature of
15000\,K, the implied mass of 0.57\,\msun\ is consistent with a
carbon-oxygen interior.  \citet{schwopemengel97} identified a narrow
emission line in EP\,Dra, which they assigned to the irradiated face
of the secondary star. The velocity amplitude
$K_2'\!=\!210\,\pm\,25$\,\kms, led to a primary mass of
$0.43\!\pm\!0.07$\,\msun. Since the narrow line was not detected over
the entire orbit, an independent confirmation is desirable. Hence,
with the possible exception of EP Dra, \hyeri\ may be the only polar
with good evidence of a low-mass primary.

A long-standing discrepancy exists between the large number of CVs
with low-mass primary stars predicted by binary population synthesis
models and the fact that none has been definitely detected so far
\citep[e.g.,][]{zorotovicetal11}. \citet{schreiberetal16} employed an
empirical version of the concept of consequential angular momentum
loss (eCAML) that enhances the standard AML caused by magnetic braking
and gravitational radiation. The considered CAML is related to nova
outbursts and affects primarily low-mass CVs, removing them
preferentially from the population. \citet{nelemansetal16} considered
asymmetric nova explosions that could provide a kick to the WD and
enhanced the mass transfer rate by the ensuing ellipticity of the
orbit. \citet{schenkeretal98} considered a Bondi-Hoyle type frictional
interaction of the secondary with the nova envelope that transfers
orbital angular momentum to the shell and may ultimately drive the
system over the stability limit. The AML adopted by
\citet{schreiberetal16} varies as $C/M_1$, with C a free
parameter. For appropriate $C$, it drives low-mass CVs into
catastrophic mass transfer and leaves CVs with massive primaries
unaffected. The frictional AML experienced by the secondary star
moving in the nova shell is proportional to the ejected mass, which
varies approximately as $R_1^4/M_1$. It is furthermore proportional to the duration
of the interaction, that is, inversely proportional to the expansion velocity
$V_\mathrm{ex}$ of the envelope. The mean expansion velocity can
exceed 1000\,\kms\ for high-mass WDs, but is much lower for nova
events on low-mass WDs, $V_\mathrm{ex}\!\simeq\!100\!-\!200$ for
0.6\,\msun\ and ``a few tens \kms '' for 0.4\,\msun\
\citep{sharaetal93}. A frictional AML that rises steeply with
decreasing mass is, therefore, a plausible proposition. The model may
explain the lack of low-mass CVs provided friction proves sufficiently
effective or an alternative mechanism as the one of
\citet{nelemansetal16} can be identified.

With a mass ratio $q\!=\!0.57\!\pm\!0.06$, \hyeri\ stays just below
the stability limit at $q\!\simeq\!0.65$
\citep{nelemansetal16,schreiberetal16}, ensuring stable mass transfer
in the absence of frictional AML. The added eCAML may render it
unstable, if sufficient angular momentum can be extracted from the
orbit. The long time scales in low-mass CVs imply, however, that
instability is delayed from the time mass transfer started, at least
by the waiting time until the first nova outburst and possibly
longer. This delay holds similarly for all nova-related CAML
descriptions. Nova outbursts in polars with a low-mass primary and
typically a rather low accretion rate are especially rare because the
ignition mass $\Delta\!M_\mathrm{ig}$ increases (i) with decreasing
$\dot M$ and (ii) with decreasing $M_1$ approximately as $R_1^4/M_1$
\citep[][their Fig.\,8]{townsleybildsten04}. For nova outbursts on a
He-core WD of 0.4\,\msun, \citet{sharaetal93} found
$\Delta\,M_\mathrm{ig}\!\simeq\!\ten{9}{-4}$\,\msun. Such a large
ignition mass implies that nova outbursts in \hyeri\ have a recurrence
time in excess of $10^7$\,yrs, given the long-term mean accretion rate
of \hyeri\ of $\ten{5}{-11}$\,\msol (see below). Hence, \hyeri\ may
have accreted already for 10 million years without having experienced
a nova outburst that lured it on to destruction. This time span
amounts to $\sim\,5$\% of the typical $\ten{2}{8}$ yrs it takes
non-magnetic CVs to reach the period gap \citep{kniggeetal11}. Hence,
most CVs with a low-mass primary, must have met their fate much
earlier than \hyeri\ if the eCAML model correctly describes the
evolution of CVs. The existence of \hyeri\ suggests that it is either
young and still doomed to destruction or is somehow peculiar, in
having managed to escape that outcome.

By its mass, the primary in \hyeri\ is consistent with a helium WD,
but residing in a close binary, it could well be a hybrid star with a
mixed He-CO composition \citep{pradamoroni09,zenatietal18}. Such WD
are created by massive mass loss of the progenitor star that
interrupts He-burning after the star has passed the tip of the
RGB. The hydrogen-deficient core mass is about 0.46\,\msun, when stars
with an initial mass $M_\mathrm{i}\!\la\!2$\,\msun\ incur the He
flash, but for initial masses of $2.2\!-\!2.6$\,\msun, He burning
starts at a core mass of $0.32\!-\!0.34$\,\msun\ and subsequent rapid
mass loss in a common envelope event could create a WD with the mass
of the primary in \hyeri\ that has a mixed He--CO composition.

With a period of 2.855\,h, \hyeri\ is nominally located in the
2.1-3.1\,h period gap \citep{kniggeetal11}, but the high-state
accretion rate $\dot{M}\!\simeq\!\ten{1.4}{-9}$\,\msol\
(Sect.~\ref{sec:sed}) demonstrates that \hyeri\ has not yet entered
its period gap, if it ever will. \citet{WickramasingheWu94} and
\citet{webbinkwickramasinghe02} argued that polars experience a lower
angular momentum loss by magnetic braking than non-magnetic CVs
because trapping of the wind from the secondary star in the WD
magnetosphere reduces the braking efficiency. They predicted that the
gap disappears for primaries with sufficiently high magnetic moments,
$\mu\!\ga\!\ten{4}{34} $\,Gcm$^3$. The WD in \hyeri\ has
$\mu\!<\!\ten{3.6}{34} $\,Gcm$^3$, based on
$R_\mathrm{wd}\!\simeq\!\ten{1.22}{9}$\,cm and a polar field strength
$B_\mathrm{p}\!\simeq\!40$\,MG of the dipole component
(Table~\ref{tab:zee}). This may or may not suffice to suppress
magnetic braking and dispose of the gap.
Even if \hyeri\ were about to enter the gap at a period of 2.85\,h,
the delayed entry compared with the standard upper edge of the gap of
3.1\,h is easily explained by a reduced level of magnetic braking.
Given, for example, a braking efficiency reduced by a factor of two
over that advocated by \citet[][their Fig.\,14, lower
panel]{kniggeetal11}, the upper edge of the gap would shift down to
2.85\,h. A low metalicity of the secondary would add to a delayed
entry into the gap \citep{webbinkwickramasinghe02}. If such a scenario
applies to \hyeri, the star may still reside above its own gap and
await entering it at some time in the future, if at all. 

The telling argument against strong magnetic braking is the low
photospheric effective temperature of the WD of \mbox{$9\!-\!10$}\,kK. The
theory of compressional heating relates \teff\ to the mean accretion
rate $\langle \dot M\rangle$ averaged over the Kelvin-Helmholtz
time scale of the non-degenerate envelope, which is of the order of
$10^6$\,yr for a low-mass WD \citep{townsleygaensicke09}. If the star
succeeds in establishing an equilibrium between heating and cooling,
its quiescent luminosity is
$L_\mathrm{q}\!=\!\ten{6}{-3}\,L_{\odot}\,\langle\dot
M\rangle_{-10}\,(M_1/M_{\odot})^{0.4}$ \citep[][their
Eq.~1]{townsleygaensicke09}, where $\langle\dot M\rangle_{-10}$ is the
accretion rate in units of $10^{-10}$\,\msol. Provided compressional
heating dominates over the congenital heat reservoir, we may equate
$L_\mathrm{q}$ to $4\pi R_1^2\sigma T_\mathrm{ph}^4$. With
$R_1\!=\!\ten{6.56}{8}(M_1/M_{\odot})^{-0.60}$ for
$M_1\!=\!0.4\!-\!0.6$\,\msun\ \citep{althausetal13}, we obtain
$T_\mathrm{ph}\!=\!16.7\langle \dot
M\rangle_{-10}^{1/4}(M_1/M_{\odot})^{0.40}$\,kK.  For
$M_1\!=\!0.42$\,\msun\ (Tab.~10), we find that $\langle\dot
M\rangle\!=\!\ten{5}{-11}$ \msol\ suffices to keep the temperature of
the WD at $9\!-\!10$\,kK. In its high state, \hyeri\ reached more
than $10^{-9}$\,\msol, implying that the long-term mean duty cycle
must be heavily weighted toward states of low accretion. The moderate
mass loss of the secondary, the reduced magnetic braking, and the
moderate inflation suggested by our best-fit seem to be in line with
the evolution of polars \mbox{as envisaged by \citet{webbinkwickramasinghe02}}.

\hyeri\ experienced a highly significant and so far unexplained change
of its orbital period by 10.5\,ms between 2011 and 2018. Similar
period variations have been observed in other post common-envelope
binaries (PCEB). Attempts to explain the observations involve either
(i) the action of additional bodies encircling the binary, causing an
apparent period variation, or (ii) solar-cycle like variations in the
internal constitution of the secondary star that change its quadrupole
moment and the gravitational pull on the primary, leading to a genuine
period variation \citep[][and references
therein]{applegate92,voelschowetal18,lanza19}. Process (i) is the
likely explanation for at least part of the period variations in
NN~Ser \citep{beuermannetal13,boursetal16}, but utterly fails in
others like QS~Vir \citep{boursetal16}, for which no stable planetary
model was found, even considering retrograde and highly inclined
orbits (Stefan Dreizler, private communication). The finding of
\citet{boursetal16} that PCEB with convective secondaries of spectral
type later than M5.5 largely lack period variations seems to favor
magnetic cycles as the driving mechanism. The mechanism of
\citet{applegate92} and the variants of \citet{voelschowetal18} and
\citet{lanza19} are appealing because they are based on physical
processes known to exist in late-type stars, but most authors agree
that they are too feeble to produce the observed amplitudes of the
period variations in many CVs \citep{voelschowetal18,lanza19}. It has
also been argued that PCEB may be a natural habitat of circumbinary
planets \citep{voelschowetal14}. The wealth of period variations
observed in PCEB and RSCVN binaries may well have more than a single
physical cause. The data presently available for \hyeri\ are not
sufficient to draw definite conclusions on the origin of the observed
period variation.

In summary, \hyeri\ is a rare example of the polar subgroup of
magnetic CVs that harbors a low-mass primary, either a helium WD or a
hybrid He-CO WD. The system may have passed through the
common-envelope phase after severe mass loss on the giant branch or
during initial He-burning \citep{pradamoroni09,zenatietal18}.
The key experiment to prove the low mass of the WD would be a direct
measurement of its radius. Unfortunately, our Sloan $griz$ photometry
of 2017 lacked the time resolution to measure the radius of the WD
from the finite ingress and egress times of the eclipse light curves
in $g$ or possibly $r$. With \hyeri\ still in a prolonged state of low
accretion as of early 2019, this task could be accomplished by
high-speed photometry at a large telescope.  If the low mass of the WD
is confirmed, a dedicated evolutionary study could establish the
origin and future evolution of \hyeri.

\begin{acknowledgements}
  We dedicate this paper to the lasting memory of Hans-Christoph
  Thomas, who analyzed part of the early data described here before
  his untimely death on 18 January 2012. We thank the anonymous
  referee for the careful reading of the paper and valuable comments
  that helped to improve this work. Part of the data were collected
  with the telescopes of the MOnitoring NEtwork of Telescopes, funded
  by the Alfried Krupp von Bohlen und Halbach Foundation, Essen, and
  operated by the Georg-August-Universit\"at G\"ottingen, the McDonald
  Observatory of the University of Texas at Austin, and the South
  African Astronomical Observatory.
  We made use of the ROSAT Data Archive of the Max-Planck-Institut
  f\"ur extraterrestrische Physik (MPE) at Garching, Germany. Part of
  the analysis is also based on observations of RE\,J0501-03 obtained
  with XMM-Newton on 2002-03-24, Obs-Id 0109460601.
  The observations at ESO in the years 2000, 2001, 2008, 2016, 2017,
  and 2018 were collected at the La Silla and Paranal sites under the
  programme IDs 66.D-0128, 66.D-0513, 082.D-0695, 098.A-9099, and
  0100.A-9099.  In establishing the spectral energy distribution in
  Fig.~\ref{fig:sed}, we
  accessed various data archives via the VizieR Photometric viewer
  operated at the CDS, Strasbourg, France
  (http://vizier.u-strasbg.fr/vizier/sed/). We quoted distances for
  \hyeri\ and EP\,Dra from the European Space Agency (ESA) mission
  Gaia (https://www.cosmos.esa.int/gaia).

\end{acknowledgements}

\bibliographystyle{aa}

\end{document}